\documentclass[
reprint,
superscriptaddress,
 amsmath,amssymb,
 aps,
]{revtex4-2}

\usepackage{graphicx}
\usepackage{subfigure}
\usepackage{setspace}
\usepackage{dcolumn}
\usepackage{bm}
\usepackage{longtable}
\usepackage{multirow}
\usepackage{tabularx}
\usepackage{ltablex}
\usepackage{booktabs}
\usepackage{xcolor}
\usepackage[normalem]{ulem}
\usepackage{lineno,hyperref}

\begin{document}

\title{Analysis of one-neutron transfer reaction in $^{18}$O + $^{76}$Se collision at 275 MeV}

\author{I. Ciraldo}
\email[]{ciraldo@lns.infn.it}
\affiliation{Istituto Nazionale di Fisica Nucleare, Laboratori Nazionali del Sud - Catania, Italy}
\affiliation{Dipartimento di Fisica e Astronomia “Ettore Majorana”, Università di Catania - Catania, Italy}

\author{F. Cappuzzello}
\affiliation{Istituto Nazionale di Fisica Nucleare, Laboratori Nazionali del Sud - Catania, Italy}
\affiliation{Dipartimento di Fisica e Astronomia “Ettore Majorana”, Università di Catania - Catania, Italy}

\author{M. Cavallaro}
\affiliation{Istituto Nazionale di Fisica Nucleare, Laboratori Nazionali del Sud - Catania, Italy}

\author{D. Carbone}
\affiliation{Istituto Nazionale di Fisica Nucleare, Laboratori Nazionali del Sud - Catania, Italy}

\author{S. Burrello}
\affiliation{Technische Universität Darmstadt, Fachbereich Physik, Institut für Kernphysik, Darmstadt, Germany}

\author{A. Spatafora}
\affiliation{Istituto Nazionale di Fisica Nucleare, Laboratori Nazionali del Sud - Catania, Italy}
\affiliation{Dipartimento di Fisica e Astronomia “Ettore Majorana”, Università di Catania - Catania, Italy}

\author{A. Gargano}
\affiliation{INFN - Sezione di Napoli, Napoli, Italy}

\author{G. De Gregorio}
\affiliation{INFN - Sezione di Napoli, Napoli, Italy}
\affiliation{Dipartimento di Matematica e Fisica, Università della Campania "Luigi Vanvitelli", Caserta, Italy}

\author{R. I. Maga\~na Vsevolodovna}
\affiliation{INFN, Sezione di Genova, Genova, Italy}

\author{L. Acosta}
\affiliation{Instituto de Fısica, Universidad Nacional Autonoma de Mexico - Mexico City, Mexico}

\author{C. Agodi}
\affiliation{Istituto Nazionale di Fisica Nucleare, Laboratori Nazionali del Sud - Catania, Italy}

\author{P. Amador-Valenzuela}
\affiliation{Instituto de Física, Universidad Nacional Autónoma de México, Mexico City, Mexico}

\author{T. Borello-Lewin}
\affiliation{Instituto de Fisica, Universidade de Sao Paulo - Sao Paulo, Brazil}

\author{G. A. Brischetto}
\affiliation{Istituto Nazionale di Fisica Nucleare, Laboratori Nazionali del Sud - Catania, Italy}
\affiliation{Dipartimento di Fisica e Astronomia “Ettore Majorana”, Università di Catania - Catania, Italy}

\author{S. Calabrese}
\affiliation{Istituto Nazionale di Fisica Nucleare, Laboratori Nazionali del Sud - Catania, Italy}
\affiliation{Dipartimento di Fisica e Astronomia “Ettore Majorana”, Università di Catania - Catania, Italy}

\author{D. Calvo}
\affiliation{Istituto Nazionale di Fisica Nucleare, Sezione di Torino, Italy}

\author{V. Capirossi}
\affiliation{Istituto Nazionale di Fisica Nucleare, Sezione di Torino, Italy}
\affiliation{DISAT, Politecnico di Torino, Italy}

\author{E. R. Chávez Lomeli}
\affiliation{Instituto de Fısica, Universidad Nacional Autonoma de Mexico - Mexico City, Mexico}

\author{M. Colonna}
\affiliation{Istituto Nazionale di Fisica Nucleare, Laboratori Nazionali del Sud - Catania, Italy}
\affiliation{Dipartimento di Fisica e Astronomia “Ettore Majorana”, Università di Catania - Catania, Italy}

\author{F. Delaunay}
\affiliation{Istituto Nazionale di Fisica Nucleare, Laboratori Nazionali del Sud - Catania, Italy}
\affiliation{Dipartimento di Fisica e Astronomia “Ettore Majorana”, Università di Catania - Catania, Italy}
\affiliation{LPC Caen, Normandie Université, ENSICAEN, UNICAEN, CNRS/IN2P3, Caen, France}

\author{H. Djapo}
\affiliation{Ankara University, Institute of Accelerator Technologies, Turkey}

\author{C. Eke}
\affiliation{Department of Mathematics and Science Education, Faculty of Education, Akdeniz University, Antalya, Turkey}

\author{P. Finocchiaro}
\affiliation{Istituto Nazionale di Fisica Nucleare, Laboratori Nazionali del Sud - Catania, Italy}

\author{S. Firat}
\affiliation{Department of Physics, Akdeniz University - Antalya, Turkey}

\author{M. Fisichella}
\affiliation{Istituto Nazionale di Fisica Nucleare, Laboratori Nazionali del Sud - Catania, Italy}

\author{A. Foti}
\affiliation{Istituto Nazionale di Fisica Nucleare, Sezione di Catania, Italy}

\author{A. Hacisalihoglu}
\affiliation{Institute of Natural Sciences, Karadeniz Teknik University - Trabzon, Turkey}

\author{F. Iazzi}
\affiliation{Istituto Nazionale di Fisica Nucleare, Sezione di Torino, Italy}
\affiliation{DISAT, Politecnico di Torino, Italy}

\author{L. La Fauci}
\affiliation{Istituto Nazionale di Fisica Nucleare, Laboratori Nazionali del Sud - Catania, Italy}
\affiliation{Dipartimento di Fisica e Astronomia “Ettore Majorana”, Università di Catania - Catania, Italy}

\author{R. Linares}
\affiliation{Instituto de Fisica, Universidade Federal Fluminense - Niteroi, Brazil}

\author{N. H. Medina}
\affiliation{Instituto de Fisica, Universidade de Sao Paulo - Sao Paulo, Brazil}

\author{M. Moralles}
\affiliation{Instituto de Pesquisas Energeticas e Nucleares IPEN/CNEN - Sao Paulo, Brazil}

\author{J. R. B. Oliveira}
\affiliation{Instituto de Fisica, Universidade de Sao Paulo - Sao Paulo, Brazil}

\author{A. Pakou}
\affiliation{Department of Physics and HINP, University of Ioannina - Ioannina, Greece}

\author{L. Pandola}
\affiliation{Istituto Nazionale di Fisica Nucleare, Laboratori Nazionali del Sud - Catania, Italy}

\author{H. Petrascu}
\affiliation{IFIN-HH - Magurele, Romania}

\author{F. Pinna}
\affiliation{Istituto Nazionale di Fisica Nucleare, Sezione di Torino, Italy}
\affiliation{DISAT, Politecnico di Torino, Italy}

\author{G. Russo}
\affiliation{Dipartimento di Fisica e Astronomia “Ettore Majorana”, Università di Catania - Catania, Italy}
\affiliation{Istituto Nazionale di Fisica Nucleare, Sezione di Catania, Italy}

\author{E. Santopinto}
\affiliation{INFN, Sezione di Genova, Genova, Italy}

\author{O. Sgouros}
\affiliation{Istituto Nazionale di Fisica Nucleare, Laboratori Nazionali del Sud - Catania, Italy}

\author{M. A. Guazzelli}
\affiliation{Centro Universitario FEI - Sao Bernardo do Campo, Brazil}

\author{S. O. Solakci}
\affiliation{Department of Physics, Akdeniz University - Antalya, Turkey}

\author{V. Soukeras}
\affiliation{Istituto Nazionale di Fisica Nucleare, Laboratori Nazionali del Sud - Catania, Italy}
\affiliation{Dipartimento di Fisica e Astronomia “Ettore Majorana”, Università di Catania - Catania, Italy}

\author{G. Souliotis}
\affiliation{Department of Chemistry and HINP, National and Kapodistrian University of Athens, Athens, Greece}

\author{D. Torresi}
\affiliation{Istituto Nazionale di Fisica Nucleare, Laboratori Nazionali del Sud - Catania, Italy}

\author{S. Tudisco}
\affiliation{Istituto Nazionale di Fisica Nucleare, Laboratori Nazionali del Sud - Catania, Italy}

\author{A. Yildirim}
\affiliation{Department of Physics, Akdeniz University - Antalya, Turkey}

\author{V. A. B. Zagatto}
\affiliation{Instituto de Fisica, Universidade Federal Fluminense - Niteroi, Brazil}

\collaboration{For the NUMEN collaboration}
\date{\today}

\begin{abstract}
\begin{description}
\item[Background]
Heavy-ion one-nucleon transfer reactions are promising tools to investigate single-particle configurations in nuclear states, with and without the excitation of the core degrees of freedom. A careful determination of the spectroscopic amplitudes of these configurations is essential for the accurate study of other direct reactions as well as beta-decays. In nucleon transfer reactions core excitations, for both target and projectile systems, are best approached via coupled channel reaction schemes. Despite being notoriously demanding in terms of computing resources, coupled channel analyses are progressively becoming more affordable even within model spaces large enough  for tackling medium mass nuclei. In this context, the $^{76}$Se($^{18}$O,$^{17}$O)$^{77}$Se reaction, here under study, gives a quantitative access to the relevant single particle orbitals and core polarization configurations built on $^{76}$Se. This is particularly relevant, since it provides data-driven information to constrain nuclear structure models for $^{76}$Se, which is the daughter nucleus in the $^{76}$Ge double beta decay. This reaction is one of the systems studied in the frame of the NUMEN project.

\item[Purpose]
We want to analyze transitions to low-lying excited states of the residual and ejectile nuclei in the $^{76}$Se($^{18}$O,$^{17}$O)$^{77}$Se one-neutron stripping reaction at 275 MeV incident energy and determine the role of single-particle and core-excitation in the description of the measured cross sections. In addition, we explore the sensitivity of the calculated cross section to different nuclear structure models.

\item[Methods]
The excitation energy spectrum and the differential cross section angular distributions are measured using the MAGNEX large acceptance magnetic spectrometer for the detection of the ejectiles and the missing mass technique for the reconstruction of the reaction kinematics. The data are compared with calculations based on distorted wave Born approximation, coupled channel Born approximation and coupled reaction channels adopting spectroscopic amplitudes for the projectile and target overlaps derived by large-scale shell model calculations and interacting boson-fermion model.

\item[Results]
Peaks in the energy spectra corresponding to groups of unresolved transitions to $^{77}$Se and $^{17}$O are identified. The experimental cross sections are extracted and compared to theoretical calculations. A remarkable agreement is found, without using any scaling factors, demonstrating that the adopted models for nuclear structure and reaction take into account the relevant aspects of the studied processes. The main transitions which contribute to the cross section of each peak are identified. 

\item[Conclusions]
The coupling with the inelastic channels feeding states in entrance and exit partitions is important in the one-neutron transfer reaction and should be accounted for in future analyses of other direct reactions such as single and double charge exchange processes involving $^{76}$Se isotope. The description of $^{77}$Se indicates the need of a large model space, in the view of an accurate description of the low lying states; a feature that should be likely accounted even for $\beta \beta$-decay studies of $^{76}$Ge.
\end{description}
\end{abstract}

\maketitle

\section{INTRODUCTION\protect\\}
\label{sec:intro}
The NUMEN (NUclear Matrix Elements for Neutrinoless double beta decay) project  \cite{Cap18,Agodi2021} proposes an innovative experimental approach toward the determination of the Nuclear Matrix Elements (NME) entering in the expression of neutrinoless double-$\beta$ decay ($0\nu\beta\beta$) half life for a large number of systems. The project is focused on the study of heavy-ion induced Double Charge Exchange (DCE) reactions, which are nuclear processes showing interesting similarities with $0\nu\beta\beta$, in particular because the initial and final nuclear states involved are the same \cite{Cappuzzello2015,Santopinto:2018nyt,Len19}. Furthermore, the simultaneous measurement of other relevant reaction channels is useful to study the competition of the direct DCE mechanism with multi-nucleon transfer processes.

DCE reactions are expected to proceed via two main mechanisms. One is characterized by the exchange of charged mesons and probes the nuclear response to first and second order isospin operators, thus being directly connected to $\beta\beta$ decay \cite{Santopinto:2018nyt,Len19,Bel20}. The other consists in the successive transfer of nucleons between the projectile and the target. Such multi-nucleon transfer component in DCE is connected to mean-field dynamics and represents an unwanted complication that needs to be precisely evaluated in DCE data analyses. Nevertheless, recent studies of specific systems have shown that its contribution to DCE cross section is negligible \cite{Ferreira2021b}. The study of multi-nucleon transfer processes in similar dynamical conditions as the explored DCE reactions is an important tool to get selective information about the involved nuclear many-body wave functions, including the mean-field dynamics and the correlations between nucleons. In this context, the exploration of one and two-nucleon transfer reactions in the $^{18}$O + $^{76}$Se collision at energies above the Coulomb barrier is particularly relevant, since the $^{76}$Se nucleus is the daughter in the $^{76}$Ge $\beta\beta$-decay and the nuclear matrix elements related to the $^{76}$Se$\rightarrow^{76}$Ge and $^{76}$Ge$\rightarrow^{76}$Se transitions are the same. Relevant information about the ground state wave functions of these nuclei is given by Schiffer et al. \cite{Sch08,Kay09}, who made precise cross sections measurements for both neutron- addition and removal reactions to determine  the occupation of valence orbits for neutrons which are  involved in the transfer or $\beta$ decay processes.

For a long time, heavy-ion multi-nucleon transfer reactions have been extensively studied \cite{Kaha77,Any74,Ful77,Oel78,Oert01,Mont14,Par15}, revealing interesting phenomena connected to single-particle, pairing correlations and cluster degrees of freedom. Among these, heavy-ion direct transfer reactions at energies above the Coulomb barrier are useful tools for obtaining precise spectroscopic information. However, in many studies, large scaling factors in the calculated cross sections were needed in order to reproduce experimental data. Nowadays the impressive progresses in computational resources has allowed a deeper insight in the application of nuclear reaction theory to data analysis, opening new opportunities to adopt heavy-ion transfer reactions as tools to investigate the nuclear structure and the reaction mechanisms \cite{Cav13,Erm16,Car17,Linares18,Car18,Ferreira2021,Cavallaro2021}.

Over the past few years, a systematic study on heavy-ion-induced one- and two-neutron transfer reactions on different target nuclei was pursued at the Istituto Nazionale di Fisica Nucleare-Laboratori Nazionali del Sud (INFN-LNS) (Italy) by the ($^{18}$O, $^{17}$O) and ($^{18}$O, $^{16}$O) reactions at incident energies ranging from 84 to 275 MeV \cite{Cav13,Erm16,Car17,Linares18,Car18,Cap15,Cav19,Car20,Paes17}. Many nuclear systems were explored using $^{9}$Be, $^{11}$B, $^{12}$C, $^{13}$C, $^{16}$O, $^{28}$Si, and $^{64}$Ni targets and the MAGNEX magnetic spectrometer \cite{Cap16,CapMag11} to detect the ejectiles. Thanks to the spectrometer high resolution and large acceptance, high quality inclusive spectra were obtained. In this perspective, $^{18}$O beam is an interesting tool to probe neutron-neutron pairing correlation in target nuclei due to the pronounced $^{16}$O core + 2n pairing configuration in its ground state. In these studies effects due to the 2n-pairing correlation have been clearly observed in the ejectiles mass distribution \cite{Agodi18}, in the spectral shapes for ($^{18}$O, $^{16}$O) reaction \cite{Cav13, Cap15,Cap12,Bon19} and in the cross section angular distributions \cite{Carbone15,Erm16,Erm17,Car18}. In addition, the exploration of ($^{18}$O, $^{17}$O) reaction has revealed an important competition of the feeding of single particle and core polarization configurations in the populated nuclear states \cite{Linares18,Car14,Cav13}. In the present context, core polarization means that the neutron is transferred onto or from an excited state, via excited core configurations.

In heavy-ion-induced transfer nuclear reactions the theoretical scenario is challenging also from the reaction mechanism point of view. It has been found that an accurate description of the distortion of the incoming and outcoming relative wave functions due to the Initial (ISI) and Final State Interactions (FSI) is mandatory. This task is well accomplished if parameter-free double folding potentials are adopted.
Distorted Wave Born Approximation (DWBA) calculations, performed within this potential models have shown a reasonable, despite not always satisfactory, description of transfer cross sections. In particular, projectile-target excitation preceding and/or following the transfer of nucleons can play a central role and, in such cases, must be taken into account properly \cite{Mer79}. A consistent way to describe such effects is to explicitly include excited states in the theoretical framework through Coupled Channel Born Approximation (CCBA) or Coupled Reaction Channel (CRC) formalisms \cite{Bond77,Lemaire77,Pereira12}. In ref. \cite{Linares18} comparisons between experimental, DWBA and CRC angle-integrated cross sections suggest that excitations before or after the transfer of a neutron are relevant in the $^{18}$O + $^{16}$O and $^{18}$O + $^{64}$Ni systems. It is also known that the effects due to core excitation both for $^{18}$O projectile and $^{76}$Se target are not negligible for the correct description of the reaction mechanism \cite{Linares18,Lemaire77,Sch08,Kay09} and influence the nucleon transfer process.

To our knowledge, no studies about the $^{76}$Se($^{18}$O,$^{17}$O)$^{77}$Se  one-neutron transfer reaction are available in literature. Although $^{18}$O was adopted as nuclear probe in the study of many reactions \cite{Re75,Bo75,Love77,Sal93,Sa01} and in others the $^{76}$Se target was used \cite{La76,Le77,Bu80}, in all of them the experimental conditions were very different from the physical case of interest for NUMEN.

Here we report, for the first time, an experimental and theoretical study of the $^{76}$Se($^{18}$O,$^{17}$O)$^{77}$Se reaction at 275 MeV incident energy. Energy spectra and cross section angular distributions for the transitions to the ground and low-lying excited states are presented. Shell-model and interacting boson fermion model calculations are performed to derive the spectroscopic amplitudes for the projectile and target overlaps. For the reaction modeling Distorted Wave Born Approximation (DWBA), Coupled Channel Born Approximation (CCBA) and coupled reaction channel (CRC) approaches are adopted. Special attention is paid to explore the role of core polarization on the populated low-lying states. This work is part of the network of reactions studied at INFN-LNS within the NUMEN and NURE \cite{Cavallaro2017} projects with the goal to extract the cross section of the $^{76}$Se($^{18}$O,$^{18}$Ne)$^{76}$Ge DCE reaction.

The experimental setup and the data reduction technique are described in Sec. \ref{sec:exp}. Section \ref{sec:teor} describes the theoretical approaches used in the data analysis and the comparison of the calculations with the experimental data. 
The obtained results are discussed in Sec. \ref{sec:discussion} and final conclusions and outlooks are given in Sec. \ref{concl}.

\section{EXPERIMENT AND DATA REDUCTION}
\label{sec:exp}
The experiment was performed at INFN-LNS in Catania where a 275 MeV $^{18}$O$^{8+}$ beam was delivered by the Superconducting Cyclotron. A thin film of $^{76}$Se (thickness 270$\pm$14 $\mu \mbox{g/cm}^{2}$) evaporated on a natural carbon backing (thickness $80\pm4$ $\mu g/cm^{2}$) was used as target. In order to estimate and subtract the contribution in the collected data from the interaction of the beam with the backing, a supplementary  measurement was performed in the same experimental conditions using a natural carbon target (thickness 400$\pm$20 $\mu \mbox{g/cm}^{2}$). The targets were produced at INFN-LNS chemical laboratory. A copper Faraday cup was used to stop the beam and measure the integrated electric charge.

\begin{figure}[hbtp]
\centering
\includegraphics[scale=0.46]{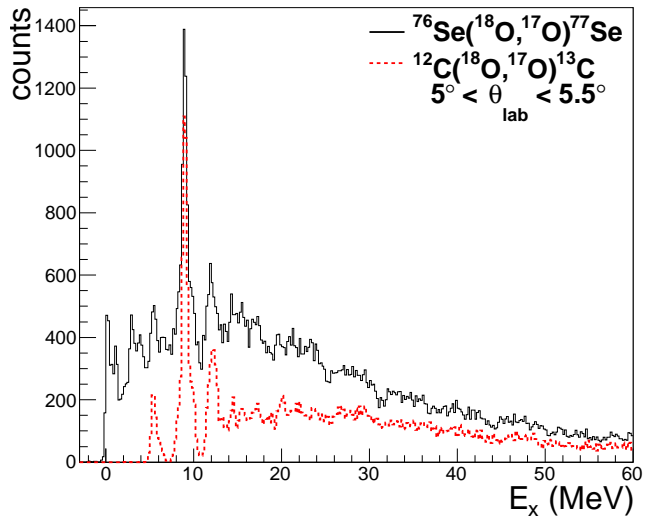}
\caption{(color online) Excitation energy spectrum for the $^{76}$Se($^{18}$O,$^{17}$O)$^{77}$Se (solid black line) and $^{12}$C($^{18}$O,$^{17}$O)$^{13}$C (dotted red line) reactions at 275 MeV incident energy in the angular range $5^\circ<\theta_{lab}<5.5^\circ$.}
\label{fig:background}
\end{figure}

The $^{17}$O$^{8+}$ reaction ejectiles were momentum analyzed by the MAGNEX magnetic spectrometer \cite{Cap16}. The parameters of the ions trajectories (i.e., vertical and horizontal positions and incident angles at the focal plane) were measured by the focal plane detector \cite{Tor21}, allowing for particle identification \cite{Cap10,Cal20}. Trajectory reconstruction of the $^{17}$O ejectiles was performed solving the equation of motion for each detected particle \cite{Cap11} to obtain scattering parameters at the target point. Further details of the data reduction technique can be found in Refs. \cite{Cap14,Carbone15,Cav11}. The Q-value and the excitation energy E$_{x}$ were obtained by missing mass calculations based on relativistic energy and momentum conservation laws: E$_{x}$ = Q$_{0}$ - Q (where Q$_{0}$ is the ground-to-ground state reaction Q-value). For the $^{76}$Se($^{18}$O,$^{17}$O)$^{77}$Se reaction, one angular setting was explored with the spectrometer optical axis centered at 8$^{\circ}$. Due to the large angular acceptance of MAGNEX, this angular setting corresponds to a total covered range of scattering angles $3^\circ<\theta_{lab}<14^\circ$.

An example of excitation energy spectrum is shown in Fig. \ref{fig:background}. The peak close to 9 MeV, originated from the reaction on carbon contaminant, was used as a reference for the background subtraction. Fig. \ref{fig:background} demonstrates that the background contribution, indicated by the red line, is not negligible at excitation energies higher than 5 MeV. Thus, only the spectrum region at lower excitation energy is studied in the present analysis.

\begin{figure}[hbtp]
\centering
\includegraphics[scale=0.55]{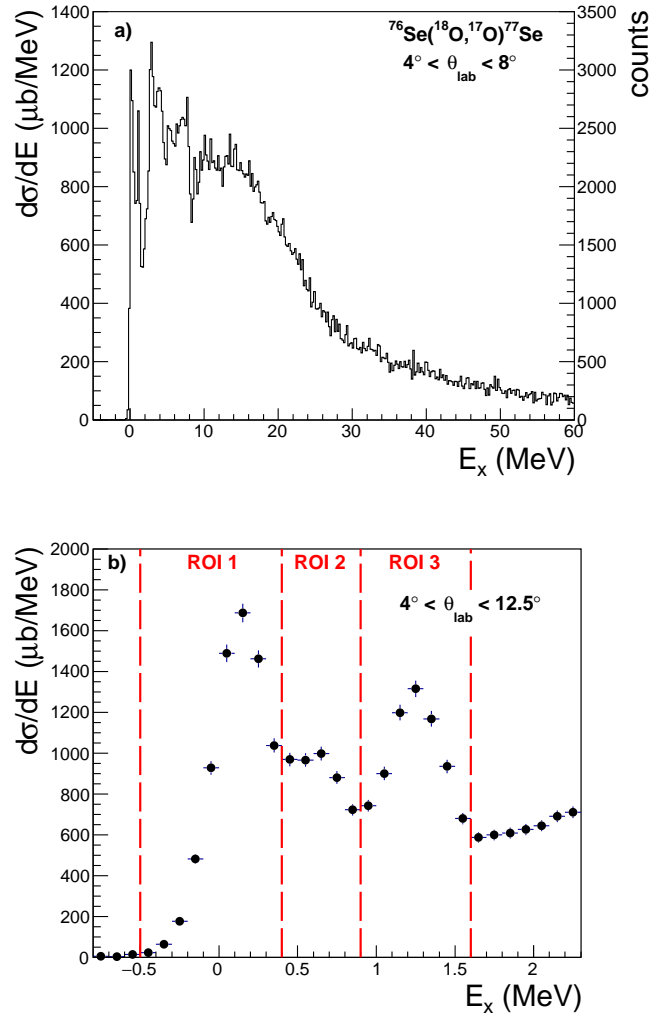} 
\caption{(a) Differential cross section spectrum obtained for the $^{76}$Se($^{18}$O,$^{17}$O)$^{77}$Se one-neutron stripping reaction at 275 MeV in the angular range $4^{\circ}<\theta_{lab}<8^{\circ}$. (b) Zoomed view at low excitation energy for $4^{\circ}<\theta_{lab}<12.5^{\circ}$. The three regions of interest (ROI) for the extraction of angular distributions are defined by the dashed red lines.}
\label{fig:xsec_fit}
\end{figure}

The absolute cross sections are extracted according to the technique described in Ref. \cite{Carbone15}, taking into account the overall MAGNEX efficiency \cite{Cav11}. The energy differential cross section spectrum for one-neutron stripping reaction, after the carbon background subtraction, is shown in Fig. \ref{fig:xsec_fit} (a) and (b). The error bars included in the spectrum indicate the statistical uncertainty. An overall uncertainty of about 10\%, not shown in the figures, is common to all the points in the spectrum, originated from the target thickness and the Faraday cup charge collection measurement. In the present experimental conditions, the achieved angular resolution is $\delta \theta _{LAB}$ (FWHM) $\sim$0.5$^{\circ}$. The energy resolution is $\delta E$ (FWHM) $\sim$310 keV. The observed structures correspond to the superposition of peaks associated to different transitions, which were not resolved due to the high level density of the residual nucleus. 

Transfer reaction cross sections between heavy ions at energies well above the Coulomb barrier are maximized around optimal values of the Q-value (Q$_{opt}$) and transferred angular momentum (L$_{opt}$) as described in \cite{Brink72}. The Q$_{opt}$ = -5 MeV estimated for the examined reaction results in the decreasing of the cross-section at increasing excitation energies as evident in the spectrum in Fig. \ref{fig:xsec_fit} (a) and, consequently, a typically larger cross section for transitions to the first low-lying states.

\begin{figure}[hbtp]
\centering
\includegraphics[scale=0.55]{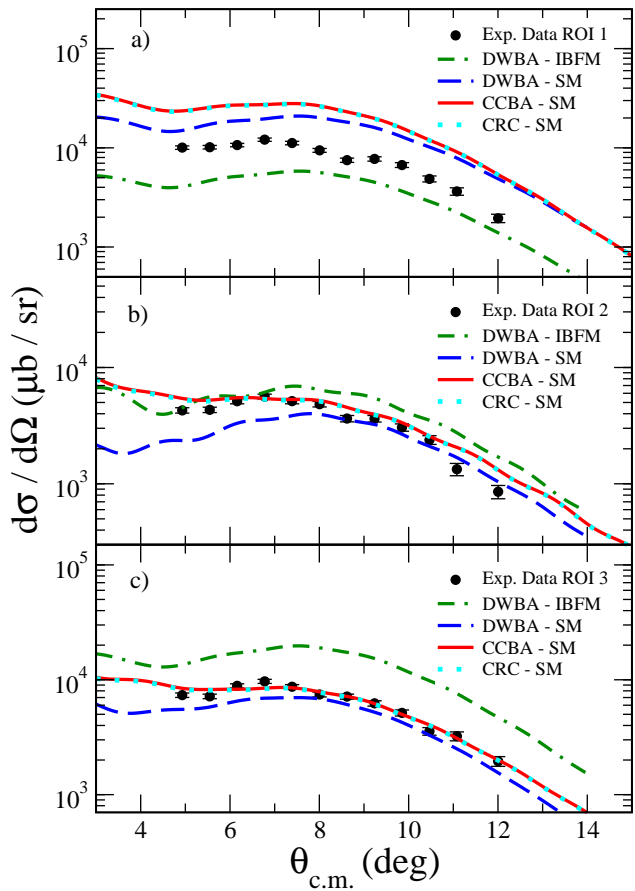}
\caption{(color online) Comparison between the theoretical curves and experimental points for one-neutron transfer angular distribution. The angular distributions related to the contribution of the unresolved excited states of the first, second and third ROI in Fig. \ref{fig:xsec_fit} (b) is shown. The  DWBA (dashed blue line), CCBA (continuous red line) and CRC (dotted cyan line) calculations obtained using shell-model spectroscopic amplitudes (SM) are shown together with DWBA (dotted-dashed green line) calculations obtained using spectroscopic amplitudes from the interacting boson-fermion model (IBFM) for the target and from shell-model for the projectile. (see text).}
\label{fig:teor_ang_distr}
\end{figure}

In Fig. \ref{fig:xsec_fit} (b) the energy differential cross section for the $^{76}$Se($^{18}$O,$^{17}$O)$^{77}$Se reaction at low excitation energy is shown. Although the limited resolution and the high level density do not allow to isolate single transitions, three main structures are visible. Three regions of interest (ROI) are considered at -0.5 $< E_{x} <$ 0.4 MeV, 0.4 $< E_{x} <$ 0.9 MeV and 0.9 $< E_{x} <$ 1.6 MeV in correspondence to the three structures.

In Fig. \ref{fig:teor_ang_distr} the experimental angular distributions for the three energy regions are plotted. Error bars include the statistical error and the uncertainties coming from the solid angle and the efficiency correction. The three extracted angular distributions are characterized by a characteristic decrease for angles larger than the grazing one ($\theta_{c.m.}\sim 9 ^{\circ}$).

\section{THEORETICAL ANALYSIS}
\label{sec:teor}

Shell-model and interacting boson fermion model calculations are performed to derive the  spectroscopic amplitudes for the projectile and target overlaps. For the reaction modeling Distorted Wave Born Approximation (DWBA), Coupled Channel Born Approximation (CCBA) and coupled reaction channel (CRC) approaches are adopted.

\subsection{Shell model calculations}
\label{sec:SA}

To obtain the spectroscopic amplitudes for both projectiles and target overlaps within the shell model framework, the KSHELL \cite{KSHELL} code was used.

For the projectile overlaps we have performed shell-model calculations adopting the Zuker-Buck-McGrory (ZBM) effective interaction  \cite{ZBM} where the $^{12}$C is considered as a closed inert core and the model space is spanned by $0p_{1/2}$, $0d_{5/2}$, and $1s_{1/2}$ orbits for both protons and neutrons.
This phenomenological potential, already adopted in many previous studies \cite{Erm16,Car17,Linares18,Cav13, Paes17,Ferreira2021}, allows to describe successfully low-lying positive as well as negative parity states for $A=16-18$. Very similar results have been obtained employing the psdmod \cite{Utsuno2011} interaction, adopted in recent works \cite{Car18,Car20,Sgouros2021}.

Regarding the target overlaps, instead, the shell-model calculations were performed considering $^{56}$Ni as closed inert core with protons and neutrons in the valence space made up of $0f_{5/2}$, $1p_{3/2}$, $1p_{1/2}$ and $0g_{9/2}$ orbitals. The effective Hamiltonian, already adopted in some recent studies \cite{Coraggio,Rocchini}, was derived within the framework of many-body perturbation theory from the CD-Bonn nucleon-nucleon potential \cite{CDBONN} renormalized by way of the V$_{\rm{low-k}}$ approach \cite{Vlowk} with the addition of the Coulomb potential for protons. In particular, the two-body matrix elements have been calculated within the $\hat Q$-box folded-diagram approach \cite{Qbox}, including in the perturbative expansion of the $\hat Q$-box one and two-body diagrams up to the third order in the interaction. 

\begin{table}[htbp]
\centering
\caption{Proton and neutron single particle energies adopted in the calculation.}
\begin{ruledtabular}
\begin{tabular}{ccc}
$\pi$(nlj) & $\epsilon_\pi$(MeV)& $\epsilon_\nu$(MeV) \\ [0.5 ex] 
 \hline
 $0f_{5/2}$ & 1.0 & 0.8  \\ 
 $1p_{3/2}$& 0.0 & 0.0  \\
$1p_{1/2}$  & 1.1 & 1.1 \\
$0g_{9/2}$ & 3.7 & 3.7 \\
\end{tabular}
\end{ruledtabular}
\label{table:p_n_en}
\end{table}

The single-neutron and single-proton energies were taken, where possible, from the experimental energy spectra of $^{57}$Ni \cite{NNDC} and $^{57}$Cu \cite{NNDC}, respectively and are reported in Table \ref{table:p_n_en}. The energy of the proton in the $0g_{9/2}$ orbital, which is not available, was chosen to be the same as that of the neutron. 
The effective charges adopted for determining the B(E2) have been calculated consistently with the Hamiltonian by employing the
Suzuki-Okamoto formalism \cite{Suzuki}. All details about the procedure can be found in \cite{Coraggio}.

The theoretical excitation energies of all the states of $^{17}$O, $^{18}$O, $^{76}$Se and $^{77}$Se nuclei involved in the calculation of the cross sections are reported in Table \ref{table:teor_en} (appendix \ref{appendix:en}) and compared with the experimental values. The calculated spectroscopic amplitudes are listed in Tables \ref{table:SA_KSHELL} (appendix \ref{appendix:ME2}). 
From Table \ref{table:teor_en}, we see that the first excited  states in $^{17}$O, $^{18}$O and $^{76}$Se are well reproduced by theory, while a less good agreement is obtained for the more complex spectrum of the odd $^{77}$Se isotope. In fact, we predict the g.s. state spin to be 9/2$^{+}$ instead of 1/2$^{-}$ and most of the excited states significantly above the experimental energies. These discrepancies may reflect the need of an enlargement of the adopted model space as discussed in Sec. \ref{sec:discussion}, as well as some inaccuracy in the matrix elements of the shell-model Hamiltonian. As a matter of fact, the employed Hamiltonian (see Ref. \cite{Rocchini}) is developed for systems with two valence particles, while in the $^{77}$Se case we are considering 21 valence nucleons.  This means that one should take into account the filling of the model space orbitals, as it was done for instance in Refs. \cite{Coraggio2020,Coraggio2021} by deriving density dependent effective Hamiltonians. Based on our previous calculations \cite{Coraggio,Rocchini} we expect the limitations of the present calculations to affect more significantly the energy spectra than the wave functions.

\subsection{Interacting boson-fermion model}
\label{sec:IBFM}
Spectroscopic amplitudes for target overlaps between $^{76}$Se and $^{77}$Se nuclei have been calculated also by using the formalism of the interacting boson model (IBM-2) and neutron-proton interacting boson-fermion model (IBFM-2), respectively.

The IBM-2 deals with even-even nuclei, where valence nucleon pairs are replaced with bosons with angular momentum 0 or 2 \cite{Iachello:2006fqa}. The IBM-2 can be extended to study odd-$A$ nuclei in the IBFM-2 by coupling an extra fermion to the boson system \cite{Iachello:2005aqa}. The IBM-2 and IBFM-2 were previously used in similar calculations in Refs. \cite{Paes17,Car20}.

\begin{table}[htbp]
\centering
\caption{IBM-2 model parameters for the $^{76}$Se  from \cite{Kaup1983,Santopinto:2018nyt}. $\epsilon_{\rm d}$, $\kappa$, $\chi_\pi$, $\chi_\nu$ and $\xi_3$ are the Hamiltonian terms related to proton and neutron boson energy, pairing, quadrupole and symmetry energy. The parameters $\xi_1$, $\xi_2$, $c_{\nu}^{(0)}$, $c_{\nu}^{(2)}$ and $c_{\nu}^{(4)}$ not shown here are set to zero.}
\begin{ruledtabular}
\begin{tabular}{cccccc}
Nucleus & $\epsilon_{\rm d}$ (MeV) & $\kappa$ (MeV)  & $\chi_\pi$ & $\chi_\nu$ &$\xi_3$ (MeV)\\
\hline
$^{76}$Se & 0.96 & $-0.16$  & $-0.9$ & 0.5 &-0.1\\
\end{tabular}
\end{ruledtabular}
\label{table:parametersIBM2}
\end{table}

\begin{table}[htbp]
\centering
\caption{Neutron single-particle energies $E_{j_\nu}$, quasi particle energies $Qspe$ and occupation probabilities $v^2$ of $^{77}$Se used in the present IBFM-2 model calculations.}
\begin{ruledtabular}
\begin{tabular}{ccccc}
Orbit $j_\nu$    &    $E_{j_\nu}$(MeV) &        $Qspe$ (MeV)   &     $v^2$\\
\hline
        $0f_{5/2}$   & 2.1980 &   1.3678  &  0.4907\\
        $0g_{9/2}$   & 1.6380 &   1.4683 &   0.6820\\
        $1p_{3/2}$   & 0.8510 &   1.9017  &  0.8474\\
        $1p_{1/2}$   & 0.0000 &   2.5670  &  0.9231\\
\end{tabular}
\end{ruledtabular}
\label{spe}
\end{table}

\begin{table}
\centering
\caption{IBFM-2 model parameters for $^{77}$Se. $A$, $\Gamma$ and $\Lambda$ are the boson-fermion Hamiltonian coefficients for describing monopole, quadrupole and exchange interaction.}
\begin{ruledtabular}
\begin{tabular}{cccc}
Nucleus & $A$ (MeV) & $\Gamma$ (MeV) & $\Lambda$ (MeV)\\
\hline
$^{77}$Se & $-1.0$ & 2.5 & 0.4\\
\end{tabular}
\end{ruledtabular}
\label{parametersIBFM2}
\end{table}

The even-even $^{76}$Se nucleus is studied in the context of the IBM-2 in Ref. \citep{Santopinto:2018nyt,Kaup1983}. The model parameters used in IBM-2 for the $^{76}$Se nucleus are fitted to reproduce its energy levels \cite{Kaup1983}. The used parameters are reported in Table \ref{table:parametersIBM2}.

$^{77}$Se is built by coupling one neutron to the core $^{76}$Se and considering the four orbitals ($0f_{5/2}$, $1p_{3/2}$, $1p_{1/2}$ and $0g_{9/2}$), as in the shell-model calculations. The odd-fermion Hamiltonian \cite{Iachello:2005aqa,Ferretti2020} and the quasi-particle energy of the odd particle are calculated in the Bardeen-Cooper-Schrieffer (BCS) approximation \cite{Bardeen:1957mv,Alonso:1984rvl,Arias:1985,Alonso:1986,Arias:1985kjc}. The neutron quasi-particle energies and occupation probabilities are calculated by solving the BCS equations and displayed in Table \ref{spe}. The unperturbed neutron single-particle energies of the $^{77}$Se isotope, $E_{j_\nu}$, required by BCS calculation are estimated by diagonalization of a Wood Saxon Potential and are also shown in Table \ref{spe}. The IBFM-2 model parameters used are reported in Table \ref{parametersIBFM2}.

A comparison between the calculated and experimental energy spectra for the $^{76}$Se and $^{77}$Se nuclei is shown in Table \ref{table:teor_en} (appendix \ref{appendix:en}) and spectroscopic amplitudes are listed in Table \ref{table:SA_KSHELL} (appendix \ref{appendix:SA}). The IBFM excitation energies are overall in a better agreement with the experimental ones as compared to shell-model predictions, signaling that more correlations are effectively accounted for in the fitting procedure performed to define the adopted Hamiltonian.

\begin{figure}[hbtp]
\centering
\includegraphics[scale=0.23]{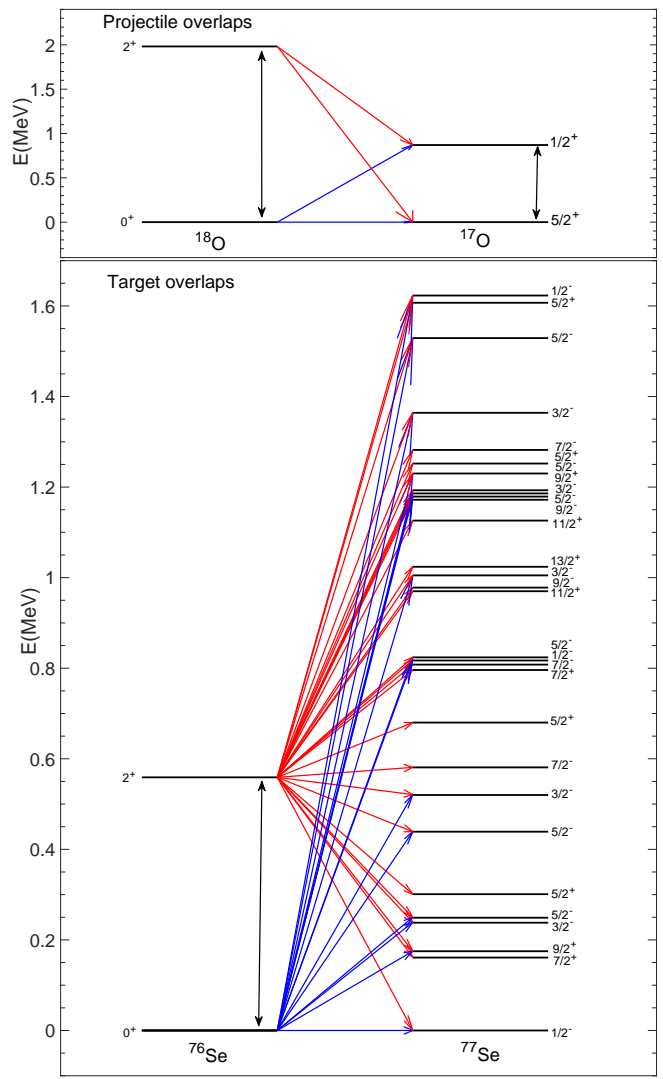}
\caption{Coupling schemes for the projectile and target overlap considered in the one-neutron transfer calculations. The blue arrows correspond to the couplings adopted in DWBA calculations.
CCBA also includes the couplings represented by red and black arrows. CRC includes the same couplings as CCBA but at infinite order. The (black) couplings between the $^{77}$Se excited states are not explicitly indicated for better readability.}
\label{fig:coupling_scheme}
\end{figure}

\subsection{Reaction calculations}
\label{sec:teor_calc}
We performed calculations using the FRESCO code \cite{Thom88,Thom09} in the $^{76}$Se($^{18}$O,$^{17}$O)$^{77}$Se one-neutron transfer reaction considering Distorted Wave Born Approximation (DWBA), Coupled Channel Born approximation (CCBA) and Coupled Reaction Channel (CRC) approaches.

The states included in the coupling scheme are sketched in Fig. \ref{fig:coupling_scheme} where the couplings between $^{77}$Se states are omitted for better readability.

The optical potentials for the ingoing and outgoing partitions were chosen according to the elastic and inelastic scattering analysis of the $^{18}$O + $^{76}$Se collision at the same energy described in Ref. \cite{LaFauci20}. The distorted waves at the entrance and exit channels were generated adopting the double-folding São Paulo potential for the description of both the real and imaginary parts of the optical potential. For the imaginary part a typical choice
is to use the same geometry as for the real one, with a scaling factor properly determined. For DWBA calculations the scaling factor is equal to 0.78 which is a standard choice \cite{Alvarez2003,Gasques06,Sousa10} for both initial and final partition. For CCBA and CRC calculations, the normalization coefficient is equal to 0.6, as typically done in order to account for all the channels not explicitly included in the system of coupled equations, like fusion and coupling to higher excitation energy bound and continuum states \cite{Pereira09}. Such prescriptions for the optical potentials have been successfully used in the analyses of several scattering, transfer and charge exchange data \cite{Carbone2021,Spat19,LaFauci20,Cap16,ZagattoPRC2018,Cavallaro2021,OliveiraNPP2013,Pereira12,Ferreira2021b,Cav13,Erm16, Car17,Linares18,Car18,Ferreira2021,Car20,Erm17,Car17,Paes17, Sgouros2021,Burrello21,Calabrese2021}. A study of the sensitivity of the results here presented to the scaling factor of the imaginary part has been performed. It shows that the cross sections are not significantly affected if the scaling factor is varied within 20\% of the standard values with a superior agreement achieved when no arbitrary variation is applied. The transfer operator was calculated in the post-representation including full complex remnant terms, as done in \cite{Sgouros2021}.
 
The single-particle wave functions are generated adopting, as core effective interactions, Woods-Saxon potentials having a reduced radius $r_{0}$ = 1.26 fm and a diffuseness $a_{0}$ = 0.70 fm \cite{Car18}. For the heavier target-like system $r_{0}$ = 1.20 fm and $a_{0}$ = 0.60 fm were adopted, as typically done when considering similar medium-mass nuclei \cite{Car20,Ferreira2021,Sgouros2021,Calabrese2021}. The depths of such Wood-Saxon potentials were adjusted in order to reproduce the experimental one-neutron separation energies.

\begin{table}[htbp]
\centering
\caption{B(E2; $0^{+} \rightarrow 2^{+}$) values for $^{18}$O and $^{76}$Se taken from experimental data (EXP) of Ref. \cite{Pritychenko16} and from shell model (SM) calculations.}
\begin{ruledtabular}
\begin{tabular}{ccc}
B(E2; $0^{+} \rightarrow 2^{+}$) & EXP ($e^{2}b^{2}$) & SM ($e^{2}b^{2}$)\\
 \hline
 $^{18}$O & 0.0043 & 0.0030  \\ 
 $^{76}$Se& 0.4320  & 0.3722 \\
\end{tabular}
\end{ruledtabular}
\label{table:be2}
\end{table}

Couplings between the states are considered both for the initial and the final partition. For the initial partition, the Coulomb and nuclear deformations for $^{18}$O and $^{76}$Se are obtained from reduced transition probabilities B(E2) taken from experimental data \cite{Pritychenko16}, as described in Ref. \cite{LaFauci20}. They are similar to those obtained from shell-model calculations, as given in Table \ref{table:be2}. The signs of M(E2) are obtained from shell-model calculations, according to the phase convention of the wave functions used to determine the spectroscopic amplitudes.
For the final partition, experimental data, when available, are often not accurate. Thus, the couplings to inelastic states are introduced through the reduced matrix elements M(E2) together with the corresponding deformation lengths $\delta_{2}$ defined in Refs. \cite{LaFauci20,Carbone2021,Cavallaro2021}, calculated by shell model and listed in Table \ref{table:ME2} (appendix \ref{appendix:ME2}). Final partition couplings are introduced only for the transitions characterized by the largest cross section. We found that the effect on the calculated cross sections of the final partition couplings is much smaller than the initial one.

In Fig. \ref{fig:teor_ang_distr} the comparison between the theoretical and experimental angular distributions for the different energy regions corresponding to the three mentioned ROIs of Fig. \ref{fig:xsec_fit} (b) is shown. No arbitrary scaling factors are used in the calculations. The DWBA, CCBA and CRC calculations obtained using shell-model spectroscopic amplitudes are shown together with the DWBA calculations obtained using spectroscopic amplitudes from the interacting boson-fermion model for the target and shell-model for the projectile.

For the sake of a direct comparison of the theoretical cross sections to the experimental data, the calculated energies are adjusted to the experimental ones and the obtained spectral distribution is folded with the experimental resolution ($\delta E$ (FWHM) $\sim$310 keV) and integrated in ROI 1,2,3.

The nuclear transitions which contribute more (see Table \ref{table:integrated_cross_sec} appendix \ref{appendix:ME2}) in ROI 1 are those feeding the $^{17}$O$_{g.s.}$ with the $^{77}$Se g.s., the first 9/2$^{+}$ state at $E_{x}$ = 0.175 MeV and the first 5/2$^{-}$ state at $E_{x}$ = 0.249 MeV. The dominant contributions in ROI 2 are given by the transition to the $^{17}$O$_{g.s.}$ with the 3/2$^{-}$(0.520 MeV) and the 5/2$^{+}$(0.680 MeV) states of $^{77}$Se and to $^{17}$O$_{0.870}$ with the $^{77}$Se g.s. In the region of the spectrum corresponding to ROI 3, the contribution of the population of various excited states is expected. The strongest channels are those populating the $^{17}$O$_{0.870}$(1/2$^{+}$) + $^{77}$Se$_{0.175}$(9/2$^{+}$), $^{17}$O$_{0.870}$(1/2$^{+}$) + $^{77}$Se$_{0.249}$(5/2$^{+}$) and $^{17}$O$_{0.870}$(1/2$^{+}$) + $^{77}$Se$_{0.520}$(3/2$^{-}$) states.

\section{DISCUSSION}
\label{sec:discussion}

Comparing in Fig. \ref{fig:teor_ang_distr} shell model DWBA and CCBA or CRC calculations, the role of the inclusion of inelastic excitations of projectile and target is evident. A global information is obtained by considering the cross section integrated in the angular range [4$\deg$ - 12.5$\deg$] of the laboratory system, which are listed in Table \ref{table:integrated_cross_sec} (appendix \ref{appendix:cross}). By summing these cross sections separately for each of the three ROIs, we get an enhancement in CCBA calculations with respect to DWBA of 30\% in ROI 1, 57\% in ROI 2 and 27\% in ROI 3. On the other hand, the curves related to CRC calculations are always practically superimposed to the ones corresponding to the CCBA results. An explanation is that the importance of the back-coupling of the transfer on the elastic channel turns out to be rather small, when focusing in the angular window we are considering. 

To gain more insight, we have analyzed in detail the angular distributions shown  in Fig. \ref{fig:teor_ang_distr}. In Fig. \ref{fig:teor_ang_distr} (a) the angular distribution obtained integrating ROI 1 is shown. Although the shell model curves are above the experimental points, the slope is very similar, especially for the CCBA and CRC calculations.
In Fig. \ref{fig:teor_ang_distr} (b) the angular distribution obtained integrating ROI 2 is shown. In this case the experimental absolute cross section is well reproduced by shell model CCBA and CRC calculations. The inclusion of the coupled channels improves the agreement with the diffraction pattern of the data. 
In Fig. \ref{fig:teor_ang_distr} (c) the angular distribution obtained integrating ROI 3 is shown. Also, in this case shell model based calculations and experimental data show an excellent agreement both in the shape and in the cross section values, especially when coupled channels are considered (CCBA and CRC).

In general, CCBA results give a larger cross section and a slightly different diffraction pattern compared to DWBA. The inclusion of inelastic excitations of projectile and target, respectively through $^{18}$O$^{2+}$ and $^{76}$Se$^{2+}$, in the coupling scheme improves the agreement with the experimental data. This is particularly true in the case of ROIs 2 and 3 where the CCBA prediction shows an excellent agreement with the experimental cross section. Although in the case of ROI 1 CCBA calculation slightly overestimates the experimental cross section, the inclusion of inelastic excitations produces a better agreement between theory and experiment. For example, the transition to the $^{17}$O$_{g.s.}$ with the the 5/2$^{+}$(0.680 MeV) state of $^{77}$Se is a dominant contribution in ROI 2.

As noticed above, calculations using shell-model spectroscopic amplitudes slightly overestimate the experimental ROI 1 cross section. This finding may indicate some inaccuracy in our predictions of the spectroscopic amplitudes.

As a test, we have compared the theoretical summed spectroscopic strengths for $^{76}$Se to $^{77}$Se with the measured ones reported in Ref. \cite{Sch08} for $l = 1,3,4$ transitions, which give information on the vacancies of the orbitals in
the target nucleus. Note that we have included in the calculations fifteen states for each angular momentum, in order to reduce the contribution of missing states to only 3\%. We predict a vacancy for the  9/2$^{+}$ state 20\% larger than that determined for  the $l = 4$ transfers, while the measured $l = 3$ strength, corresponding essentially to transitions to 5/2$^{-}$ states, is quite well reproduced. On the other hand, theory underestimates the measured $l = 1$ strength (0.70 versus 1.63), which results from the combination of transitions to the 1/2$^{-}$ and 3/2$^{-}$ states, and therefore cannot provide separate information for each angular momentum. In other words, it turns out that within our shell model calculations 1.3 neutrons are missed in the occupation of the 0g$_{9/2}$ orbital of the $^{76}$Se$_{gs}$, which are, instead, allocated in the 1p$_{3/2}$ and 1p$_{1/2}$ orbitals.

The above comparison, as well as the discrepancies we find between the experimental and theoretical excitation energies (see Table \ref{table:teor_en} appendix \ref{appendix:en}), may evidence the incompleteness of our adopted model space, which does not include the neutron 1d$_{5/2}$ and the proton 0f$_{7/2}$ orbitals. No shell-model calculations are available for Se isotopes in such a large model space. However, clear indications about the role of excitations outside the 0f$_{5/2}$ 1p 0g$_{9/2}$ model space are available for Zn isotopes \cite{Wraith2017,Xie2019,Louchart2013}, with only 4 protons less with respect to Se.

From an extension of our model space, we should expect a decrease in the vacancy of the 0g$_{9/2}$ orbital inducing a reduction in the related spectroscopic amplitudes and, therefore, a reduction of the theoretical cross section.

It is worth mention that the importance of the neutron 1d$_{5/2}$ orbital has been shown also to reproduce the spectroscopy of N $\geq$ 40 neutron-rich nuclei in the region of heavy Fe (Z=26) and Cr (Z=24) isotopes (see, for instance,  Ref. \cite{Lenzi2010}).

In order to probe the cross section sensitivity to different theoretical models of nuclear structure when the reaction mechanism is set, we have performed exploratory calculations. In particular, we have calculated the DWBA cross sections by using spectroscopic amplitudes from IBFM for the target and from shell model for the projectile. It was not possible to perform CCBA calculations with the IBFM because spectroscopic amplitudes have been calculated (see Table \ref{table:SA_KSHELL} appendix \ref{appendix:SA}) only for states coming from the 0$^{+}$ boson coupled to the the five orbitals $0f_{5/2}$, $1p_{3/2}$, $1p_{1/2}$ and $0g_{9/2}$ orbitals, and not for states arising from the coupling with the 2$^{+}$ boson. The complete calculation will be the subject of another paper in preparation. 
As can be seen in Fig. \ref{fig:teor_ang_distr}, the experimental cross section, at variance with shell model results, is underestimated by the calculations using IBFM in the ROI 1, while in the ROIs 2 and 3 is overestimated.

These differences between shell model and IBFM results can be understood by comparing the corresponding
spectroscopic amplitudes for $^{76}$Se$_{g.s.} (0^+) \rightarrow ^{77}$Se transitions listed in Table \ref{table:SA_KSHELL} (appendix \ref{appendix:SA}). The strength is distributed between the two 9/2$^{+}$ states for IBFM whereas it is almost entirely attributed to the first 9/2$^{+}$ state for shell model. Similar distributions can be noticed in the case of 1/2$^{-}$, 3/2$^{-}$ and 5/2$^{-}$ states. Although a general rule could not be found, it seems that IBFM model distributes almost equally the strength among all the states of the same orbital whereas shell model almost entirely attributes the strength to only one of these states.

\section{CONCLUSIONS}
\label{concl}
In the present work, the $^{76}$Se($^{18}$O,$^{17}$O)$^{77}$Se one-neutron stripping reaction at 275 MeV incident energy was studied for the first time. The experiment was performed at the INFN-LNS laboratory in Catania in the context of the NUMEN experimental campaign.\\
Energy spectra and cross section angular distributions for transitions to low-lying states were extracted. Due to the high level density for the residual odd nucleus and the finite energy resolution, it was not possible to disentangle transitions to isolated states but angular distributions for transitions to group of states were explored.\\
The experimental angular distributions were compared to theoretical calculations based on DWBA, CCBA and CRC approaches. The optical potentials for the ingoing partition was chosen according to the elastic and inelastic scattering analysis of the $^{18}$O + $^{76}$Se collision. 
The one-neutron spectroscopic amplitudes for the projectile and target overlaps were derived by shell-model and interacting boson-fermion model calculations with the 0f$_{5/2}$ 1p 0g$_{9/2}$ model space. An overall good agreement with the experimental cross sections is obtained for the analyzed transitions considering that no arbitrary scaling factors have been used in the calculations. For the shell model results the inclusion of the 1d$_{5/2}$ neutron and 0f$_{7/2}$ proton orbitals in the adopted model space is envisaged for an improved description of the transitions to ${9/2}^{+}$ and ${5/2}^{-}$ states of $^{77}$Se. \\
From the comparison between the experimental cross sections and the theoretical ones calculated with shell-model spectroscopic amplitudes, it emerges that discrepancies observed in ROI 1 may be related to the incompleteness of the adopted model space. Indeed, it does not include Z = 28 and N = 50 cross shell excitations. By enlarging the model space, we expect an increase in the predicted occupation number of the 0g$_{9/2}$ orbital and a depletion of the a 1p orbitals, that would improve the agreement with the experimental data. Actually, within the present calculations, the occupation numbers of these single-particle states are, respectively, underestimated and overestimated by ~20\% respect to the experimental values reported in Ref. \cite{Sch08}.\\
We found that all the adopted models allow for a reasonable description of the data, supporting the validity of the nuclear structure and optical model inputs. However, a better agreement is obtained when the coupling to inelastic vibrational states is explicitly introduced in coupled channel calculations. This suggests a relevant role of core polarization in the $^{17}$O and $^{77}$Se odd nuclei. Instead, a negligible effect in the results was found when considering the CRC approach with respect to the CCBA one indicating a minor role for the coupling to the scattering channels of the transfer ones.\\
We expect that the mentioned couplings to inelastic states are likely to play a role in all quasielastic reactions originating from $^{18}$O + $^{76}$Se collision, including single and double charge exchange reactions of interest for NUMEN. Coupled channel analysis are thus envisaged for this purpose.

\section*{Acknowledgments}
This project has received funding from the European Research Council (ERC) under the European Union's Horizon 2020 research and innovation programme (Grant Agreement No. 714625).

Support from the Brazilian funding agencies FAPESP Proc. No. 2019/07767-1 and INCT-FNA Proc. No. 464898/2014-5 is acknowledged.

The Mexican authors acknowledge financial support
from CONACyT 314857 and DGAPA-PAPIIT IN107820 and
IG101120.

S. B. acknowledges support from the Alexander von Humboldt foundation.

We also acknowledge the CINECA award under the ISCRA initiative (code HP10B51E4M) and through the INFN-CINECA agreement for the availability of high performance computing resources and support.

G.D.G. acknowledges the support by the funding program VALERE of Univerità degli Studi della Campania “Luigi
Vanvitelli.”

\clearpage

\bibliography{aipsamp}

\clearpage

\begin{thebibliography}{99}
\bibitem{Cap18}
 F. Cappuzzello, C. Agodi, M. Cavallaro et al., Eur. Phys. J. A54, 72 (2018).
\bibitem{Agodi2021}
C. Agodi et al., Universe 7, 72 (2021).
\bibitem{Cappuzzello2015}
F. Cappuzzello, M. Cavallaro, C. Agodi et al., Eur. Phys. J. A 51, 145 (2015).
\bibitem{Santopinto:2018nyt}
E. Santopinto, H. García-Tecocoatzi, R. I. Magana Vsevolodovna, and J. Ferretti, Phys. Rev. C 98, 061601(R) (2018).
\bibitem{Len19}
H. Lenske, F. Cappuzzello, M. Cavallaro, and M. Colonna, Prog. Part. Nucl. Phys. 109, 103716 (2019).
\bibitem{Bel20}
J. Bellone et al., Phys. Lett. B 807, 135528 (2020).
\bibitem{Ferreira2021b}
J. L. Ferreira, J. Lubian, F. Cappuzzello, M. Cavallaro, and D. Carbone, Phys. Rev. C 105, 014630 (2022).
\bibitem{Sch08}
J. P. Schiffer, S. J. Freeman, J. A. Clark, C. Deibel, C. R. Fitzpatrick, S. Gros, A. Heinz, D. Hirata, C. L. Jiang, B. P. Kay, A. Parikh, P. D. Parker, K. E. Rehm, A. C. C. Villari, V.Werner, and C. Wrede, Phys. Rev. Lett. 100, 112501 (2008).
\bibitem{Kay09}
B. P. Kay et al., Phys. Rev. C 79, 021301(R) (2009).
\bibitem{Kaha77}
S. Kahana and A. J. Baltz, One- and two-nucleon transfer reactions
with heavy ions, in Advances in Nuclear Physics, edited by M. Baranger and E. Vogt, Advances in Nuclear Physics, Vol. 9 (Springer, Boston, MA, 1977).
\bibitem{Any74}
N. Anyas-Weiss et al., Phys. Rep. 12, 201 (1974).
\bibitem{Ful77}
H. Fulbright, C. L. Bennett, R. A. Lindgren, R. G.Markham, S. C. McGuire, G. C. Morrison, U. Strohbusch, and J. Toke, Nucl. Phys. A 284, 329 (1977).
\bibitem{Oel78}
W. Oelert, A. Djaloeis, C. Mayer-Böricke, P. Turek, and S. Wiktor, Nucl. Phys. A 306, 1 (1978).
\bibitem{Oert01}
W. von Oertzen and A. Vitturi, Rep. Prog. Phys. 64, 1247 (2001).
\bibitem{Mont14}
D. Montanari, L. Corradi, S. Szilner, G. Pollarolo, E. Fioretto, G. Montagnoli, F. Scarlassara, A. M. Stefanini, S. Courtin, A. Goasduff, F. Haas, D. Jelavic Malenica, C. Michelagnoli, T. Mijatovic, N. Soic, C. A. Ur, and M. Varga Pajtler, Phys. Rev.Lett. 113, 052501 (2014).
\bibitem{Par15}
A. Parmar, Sonika, B. J. Roy, V. Jha, U. K. Pal, T. Sinha, S. K. Pandit, V. V. Parkar, K. Ramachandran, K. Mahata et al., Nucl. Phys. A 940, 167 (2015).
\bibitem{Cav13}
M. Cavallaro, F. Cappuzzello, M. Bondi, D. Carbone, V. N. Garcia, A. Gargano, S. M. Lenzi, J. Lubian, C. Agodi, F. Azaiez, M. DeNapoli, A. Foti, S. Franchoo, R. Linares, D. Nicolosi, M. Niikura, J. A. Scarpaci, and S. Tropea, Phys. Rev. C 88, 054601 (2013).
\bibitem{Erm16}
M. J. Ermamatov, F. Cappuzzello, J. Lubian, M. Cubero, C. Agodi, D. Carbone, M. Cavallaro, J. L. Ferreira, A. Foti, V. N. Garcia, A. Gargano, J. A. Lay, S. M. Lenzi, R. Linares, G. Santagati, and A. Vitturi, Phys. Rev. C 94, 024610 (2016).
\bibitem{Car17}
D. Carbone, J. L. Ferreira, F. Cappuzzello, J. Lubian, C. Agodi, M. Cavallaro, A. Foti, A. Gargano, S. M. Lenzi, R. Linares, and G. Santagati, Phys. Rev. C 95, 034603 (2017).
\bibitem{Linares18}
R. Linares, M. J. Ermamatov, J. Lubian, F. Cappuzzello, D. Carbone, E. N. Cardozo, M. Cavallaro, J. L. Ferreira, A. Foti, A. Gargano, B. Paes, G. Santagati, and V. A. B. Zagatto, Phys. Rev. C 98, 054615 (2018).
\bibitem{Car18}
E. N. Cardozo, J. Lubian, R. Linares, F. Cappuzzello, D. Carbone, M. Cavallaro, J. L. Ferreira, A. Gargano, B. Paes, and G. Santagati, Phys. Rev. C 97, 064611 (2018).
\bibitem{Ferreira2021}
J. L. Ferreira, D. Carbone, M. Cavallaro, N. N. Deshmukh, C. Agodi, G. A. Brischetto, S. Calabrese, F. Cappuzzello, E. N. Cardozo, I. Ciraldo, M. Cutuli, M. Fisichella, A. Foti, L. LaFauci, O. Sgouros, V. Soukeras, A. Spatafora, D. Torresi, and J. Lubian, Phys. Rev. C 103, 054604 (2021).
\bibitem{Cavallaro2021}
M. Cavallaro et al., Front. Astron. Space Sci. 8, 659815 (2021).
\bibitem{Cap15}
F. Cappuzzello, D. Carbone,M. Cavallaro,M. Bondì, C. Agodi, F. Azaiez, A. Bonaccorso, A. Cunsolo, L. Fortunato, A. Foti et al., Nat. Commun. 6, 6743 (2015).
\bibitem{Cav19}
M. Cavallaro, F. Cappuzzello, D. Carbone, and C. Agodi, Eur. Phys. J. A 55, 244 (2019).
\bibitem{Car20}
D. Carbone, J. L. Ferreira, S. Calabrese, F. Cappuzzello, M. Cavallaro, A. Hacisalihoglu, H. Lenske, J. Lubian, R. I. Magnana Vsevolodovna, E. Santopinto, C. Agodi, L. Acosta, D. Bonanno, T. Borello-Lewin, I. Boztosun, G. A. Brischetto, S. Burrello, D. Calvo, E. R. Chavez Lomeli, I. Ciraldo, M. Colonna, F. Delaunay, N. Deshmukh, P. Finocchiaro, M. Fisichella, A. Foti, G. Gallo, F. Iazzi, L. LaFauci, G. Lanzalone, R. Linares, N. H. Medina, M. Moralles, J. R. B. Oliveira, A. Pakou, L. Pandola, H. Petrascu, F. Pinna, S. Reito, G. Russo, O. Sgouros, S. O. Solakci, V. Soukeras, G. Souliotis, A. Spatafora, D. Torresi, S. Tudisco, A. Yildirin, and V. A. B. Zagatto, Phys. Rev. C 102, 044606 (2020).
\bibitem{Paes17}
B. Paes, G. Santagati, R. M. Vsevolodovna, F. Cappuzzello, D. Carbone, E. N. Cardozo, M. Cavallaro, H. Garcia-Tecocoatzi, A. Gargano, J. L. Ferreira, S. M. Lenzi, R. Linares, E. Santopinto, A. Vitturi, and J. Lubian, Phys. Rev. C 96, 044612 (2017).
\bibitem{Cap16}
F. Cappuzzello, C. Agodi, D. Carbone, and M. Cavallaro, Eur. Phys. J. A 52, 169 (2016).
\bibitem{CapMag11}
F. Cappuzzello, D. Carbone, M. Cavallaro, and A. Cunsolo, in Magnets: Types, Uses, and Safety (Nova Science Publishers, New York, 2011), pp. 1–63.
\bibitem{Agodi18}
C. Agodi, G. Giuliani, F. Cappuzzello, A. Bonasera, D. Carbone, M. Cavallaro, A. Foti, R. Linares, and G. Santagati, Phys. Rev. C 97, 034616 (2018).
\bibitem{Cap12}
F. Cappuzzello et al., Phys. Lett. B 711, 347 (2012).
\bibitem{Bon19}
A. Bonaccorso, F. Cappuzzello, D. Carbone, M. Cavallaro, G. Hupin, P. Navratil, and S. Quaglioni, Phys. Rev. C 100, 024617 (2019).
\bibitem{Carbone15}
D. Carbone, Eur. Phys. J. Plus 130, 143 (2015).
\bibitem{Erm17}
M. J. Ermamatov et al., Phys. Rev. C 96, 044603 (2017).
\bibitem{Car14}
D. Carbone, M. Bondì, A. Bonaccorso, C. Agodi, F. Cappuzzello, M. Cavallaro, R. J. Charity, A. Cunsolo, M. De Napoli, and A. Foti, Phys. Rev. C 90, 064621 (2014).
\bibitem{Mer79}
M. C. Mermaz, A. Greiner, B. T. Kim, M. A. G. Fernandes, N. Lisbona, E.Muller,W. Chung, and B. H.Wildenthal, Phys. Rev. C 20, 2130 (1979).
\bibitem{Bond77}
P. D. Bond, H. J. Korner,M.-C. Lemaire, D. J. Pisano, and C. E. Thorn, Phys. Rev. C 16, 177 (1977).
\bibitem{Lemaire77}
M.-C. Lemaire and K. S. Low, Phys. Rev. C 16, 183 (1977).
\bibitem{Pereira12}
D. Pereira et al., Phys. Lett. B 710, 426 (2012).
\bibitem{Re75}
K. E. Rehm et al., Phys. Rev. C 12, 1945 (1975).
\bibitem{Bo75}
H. G. Bohlen et al., Z. Phys. A 273, 211 (1975).
\bibitem{Love77}
W. G. Love, T. Terasawa, and G. R. Satchler, Nucl. Phys. A 291, 183 (1977).
\bibitem{Sal93}
S. Salem-Vasconcelos, E. M. Takagui, M. J. Bechara, K. Koide, O. Dietzsch, A. B. Nuevo, and H. Takai, Phys. Rev. C 50, 927 (1994).
\bibitem{Sa01}
P. K. Sahu, R. K. Choudhury, D. C. Biswas, and B. K. Nayak, Phys. Rev. C 64, 014609 (2001).
\bibitem{La76}
J. Lachkar et al., Phys. Rev. C 14, 933 (1976).
\bibitem{Le77}
R. Lecomte, P. Paradis, J. Barrette, M. Barrette, G. Lamoureux, and S. Monaro, Nucl. Phys. A 284, 123 (1977).
\bibitem{Bu80}
H. R. Burgl et al., Nucl. Phys. A 334, 413 (1980).
\bibitem{Cavallaro2017}
M. Cavallaro et al., PoS (BORMIO2017), 015 (2017).
\bibitem{Tor21}
D. Torresi et al., Nucl. Instrum. Methods Phys. Res., Sect. A 989, 164918 (2021).
\bibitem{Cap10}
F. Cappuzzello et al., Nucl. Instrum. Methods Phys. Res., Sect. A 621, 419 (2010).
\bibitem{Cal20}
S. Calabrese et al., Nucl. Instrum. Methods Phys. Res., Sect. A 980, 164500 (2020).
\bibitem{Cap11}
F. Cappuzzello, D. Carbone, and M. Cavallaro, Nucl. Instrum. Methods Phys. Res., Sect. A 638, 74 (2011).
\bibitem{Cap14}
F. Cappuzzello, C. Agodi, M. Bondì, D. Carbone, M. Cavallaro, A. Cunsolo, M. D. Napoli, A. Foti, and D. Nicolosi, Nucl. Instrum. Methods Phys. Res., Sect. A 763, 314 (2014).
\bibitem{Cav11}
M. Cavallaro, F. Cappuzzello, D. Carbone, A. Cunsolo, A. Foti, and R. Linares, Nucl. Instrum. Methods Phys. Res., Sect. A 637,
77 (2011).
\bibitem{Brink72}
D. M. Brink, Phys. Lett. B 40, 37 (1972).
\bibitem{KSHELL}
N. Shimizu, T. Mizusaki, T. Utsuno, and Y. Tsunoda, Comput. Phys. Commun. 244, 372 (2019).
\bibitem{ZBM}
A. P. Zuker, B. Buck, and J. B. McGrory, Phys. Rev. Lett. 21, 39 (1968).
\bibitem{Utsuno2011}
Y. Utsuno and S. Chiba, Phys. Rev. C 83, 021301(R) (2011).
\bibitem{Sgouros2021}
O. Sgouros, M. Cavallaro, F. Cappuzzello, D. Carbone, C. Agodi, A. Gargano, G. De Gregorio, C. Altana, G. A. Brischetto, S. Burrello, S. Calabrese, D. Calvo, V. Capirossi, E. R. Chavez Lomeli, I. Ciraldo, M. Cutuli, F. Delaunay, H. Djapo, C. Eke, P. Finocchiaro, M. Fisichella, A. Foti, A. Hacisalihoglu, F. Iazzi, L. LaFauci, R. Linares, J. Lubian, N. H.
Medina,M.Moralles, J. R. B. Oliveira, A. Pakou, L. Pandola, F. Pinna, G. Russo, M. A. Guazzelli, V. Soukeras, G. Souliotis, A. Spatafora, D. Torresi, A. Yildirim, and V. A. B. Zagatto, Phys. Rev. C 104, 034617 (2021).
\bibitem{Coraggio}
L. Coraggio, L. De Angelis, T. Fukui, A. Gargano, N. Itaco, and F. Nowacki, Phys. Rev. C 100, 014316 (2019).
\bibitem{Rocchini}
M. Rocchini, K. Hadynska-Klek, A. Nannini, A. Goasduff, M. Zielinska, D. Testov, T. R. Rodriguez, A. Gargano, F. Nowacki, G. DeGregorio, H. Naidja, P. Sona, J. J. Valiente-Dobon, D. Mengoni, P. R. John, D. Bazzacco, G. Benzoni, A. Boso, P. Cocconi, M. Chiari, D. T. Doherty, F. Galtarossa, G. Jaworski, M. Komorowska, N. Marchini, M. Matejska-Minda, B. Melon, R. Menegazzo, P. J. Napiorkowski, D. Napoli, M. Ottanelli, A. Perego, L. Ramina, M. Rampazzo, F. Recchia, S. Riccetto, D. Rosso, and M. Siciliano, Phys. Rev. C 103, 014311 (2021).
\bibitem{CDBONN}
R. Machleidt, Phys. Rev. C 63, 024001 (2001).
\bibitem{Vlowk}
S. Bonger, T. T. S. Kuo, and L. Coraggio, Nucl. Phys. A 684, 432 (2001).
\bibitem{Qbox}
L. Coraggio, A. Covello, A. Gargano et al., Ann. Phys. 327, 2125 (2012).
\bibitem{NNDC}
Nndc, Data extracted using the NNDC On-Line Data Service
from the ENSDF database, http://www.nndc.bnl.gov/ensdf/.
\bibitem{Suzuki}
K. Suzuki and R. Okamoto, Prog. Theor. Phys. 93, 905 (1995).
\bibitem{Coraggio2020}
L. Coraggio, G. De Gregorio, A. Gargano, N. Itaco, T. Fukui, Y. Z. Ma, and F. R. Xu, Phys. Rev. C 102, 054326 (2020).
\bibitem{Coraggio2021}
L. Coraggio, G. De Gregorio, A. Gargano, N. Itaco, T. Fukui,
Y. Z. Ma, and F. R. Xu, Phys. Rev. C 104, 054304 (2021).
\bibitem{Iachello:2006fqa}
F. Iachello and A. Arima, Adv. Nucl. Phys. 13, 139 (1984).
\bibitem{Iachello:2005aqa}
F. Iachello and P. V. Isacker, The Interacting Boson-Fermion Model, Cambridge Monographs on Mathematical Physics (Cambridge University Press, New York, 1991)
\bibitem{Kaup1983}
U. Kaup, C. Mönkemeyer, and P. v. Brentano, Z. Phys. A 310, 129 (1983).
\bibitem{Ferretti2020}
J. Ferretti, J. Kotila, R. I. Magana Vsevolodovna, and E. Santopinto, Phys. Rev. C 102, 054329 (2020).
\bibitem{Bardeen:1957mv}
J. Bardeen, L. N. Cooper, and J. R. Schrieffer, Phys. Rev. 108, 1175 (1957).
\bibitem{Alonso:1984rvl}
C. E. Alonso, J. M. Arias, R. Bijker, and F. Iachello, Phys. Lett. B 144, 141 (1984).
\bibitem{Arias:1985}
J. M. Arias, PhD dissertation, University of Sevilla (1985).
\bibitem{Alonso:1986}
C. E. Alonso, PhD dissertation, University of Sevilla (1986).
\bibitem{Arias:1985kjc}
J. M. Aria, C. E. Alonso, and R. Bijker, Nucl. Phys. A 445, 333 (1985).
\bibitem{Thom88}
I. J. Thompson, Comp. Phys. Rep. 7, 167 (1988).
\bibitem{Thom09}
I. Thompson and F. Nunes, Nuclear Reactions for Astrophysics: Principles, Calculation and Applications of Low-Energy Reactions, 1st ed. (Cambridge University Press, Cambridge, UK, 2009).
\bibitem{LaFauci20}
L. La Fauci, A. Spatafora, F. Cappuzzello, C. Agodi, D. Carbone, M. Cavallaro, J. Lubian, L. Acosta, P. Amador-Valenzuela, T. Borello-Lewin, G. A. Brischetto, S. Calabrese, D. Calvo, V. Capirossi, E. R. Chavez Lomeli, I. Ciraldo, M. Cutuli, F. Delaunay, H. Djapo, C. Eke, P. Finocchiaro, S. Firat, M. Fisichella, A. Foti, M. A. Guazzelli, A. Hacisalihoglu, F. Iazzi, R. Linares, J. Ma, N. H. Medina, M. Moralles, J. R. B. Oliveira, A. Pakou, L. Pandola, H. Petrascu, F. Pinna, P. C. Ries, G. Russo, O. Sgouros, S. O. Solakci, V. Soukeras, G. Souliotis, D. Torresi, S. Tudisco, J. Wang, Y. Yang, A. Yildirin, and V. A. B. Zagatto, Phys. Rev. C 104, 054610 (2021).
\bibitem{Alvarez2003}
M. A. G. Alvarez et al., Nucl. Phys. A 723, 93 (2003).
\bibitem{Gasques06}
L. R. Gasques, L. C. Chamon, P. R. S. Gomes, and J. Lubian, Nucl. Phys. A 764, 135 (2006).
\bibitem{Sousa10}
D. P. Sousa, D. Pereira, J. Lubian, L. C. Chamon, J. R. B. Oliveira, E. S. R., Jr., C. P. Silva, P. N. de Faria, V. GuimarÃ£es, R. Lichtenthaler, and M. A. G. Alvarez, Nucl. Phys. A 836, 1 (2010).
\bibitem{Pereira09}
D. Pereira, J. Lubian, J. R. B. Oliveira, D. P. de Souza, and L. C. Chamon, Phys. Lett. B 670, 330 (2009).
\bibitem{Carbone2021}
D. Carbone et al., Universe 7, 58 (2021).
\bibitem{Spat19}
A. Spatafora et al., Phys. Rev. C 100, 034620 (2019).
\bibitem{ZagattoPRC2018}
V. A. B. Zagatto, F. Cappuzzello, J. Lubian, M. Cavallaro, R. Linares, D. Carbone, C. Agodi, A. Foti, S. Tudisco, J. S. Wang, J. R. B. Oliveira, and M. S. Hussein, Phys. Rev. C 97, 054608 (2018).
\bibitem{OliveiraNPP2013}
J. R. B. Oliveira et al., J. Phys. G: Nucl. Part. Phys. 40, 105101 (2013).
\bibitem{Burrello21}
S. Burrello, S. Calabrese, F. Cappuzzello, D. Carbone, M. Cavallaro, M. Colonna, J. A. Lay, H. Lenske, C. Agodi, J. L. Ferreira, S. Firat, A. Hacisalihoglu, L. LaFauci, A. Spatafora, L. Acosta, J. I. Bellone, T. Borello-Lewin, I. Boztosun, G. A. Brischetto, D. Calvo, E. R. Chavez-Lomeli, I. Ciraldo, M. Cutuli, F. Delaunay, P. Finocchiaro, M. Fisichella, A. Foti, F. Iazzi, G. Lanzalone, R. Linares, J. Lubian,M.Moralles, J. R. B.
Oliveira, A. Pakou, L. Pandola, H. Petrascu, F. Pinna, G. Russo, O. Sgouros, S. O. Solakci, V. Soukeras, G. Souliotis, D. Torresi, S. Tudisco, A. Yildirin, and V. A. B. Zagatto, Phys. Rev. C 105, 024616 (2022).
\bibitem{Calabrese2021}
S. Calabrese, M. Cavallaro, D. Carbone, F. Cappuzzello, C. Agodi, S. Burrello, G. DeGregorio, J. L. Ferreira, A. Gargano, O. Sgouros, L. Acosta, P. Amador-Valenzuela, J. I. Bellone, T. Borello-Lewin, G. A. Brischetto, D. Calvo, V. Capirossi, E. R. Chavez Lomeli, I. Ciraldo, M. Colonna, F. Delaunay, H. Djapo, C. Eke, P. Finocchiaro, S. Firat, M. Fisichella, A. Foti, M. A. Guazzelli, A. Hacisalihoglu, F. Iazzi, L. LaFauci, J. A. Lay, R. Linares, J. Lubian, N. H.Medina, M. Moralles, J. R. B. Oliveira, A. Pakou, L. Pandola, H. Petrascu, F. Pinna, G. Russo, S. O. Solakci, V. Soukeras, G. Souliotis, A. Spatafora, D. Torresi, S. Tudisco, A. Yildirim, and V. A. B. Zagatto, Phys. Rev. C 104, 064609 (2021).
\bibitem{Pritychenko16}
B. Pritychenko, M. Birch, B. Singh, and M. Horoi, At. Data Nucl. Data Tables 107, 1 (2016).
\bibitem{Wraith2017}
C. Wraith et al., Phys. Lett. B 771, 385 (2017).
\bibitem{Xie2019}
L. Xie et al., Phys. Lett. B 797, 134805 (2019).
\bibitem{Louchart2013}
C. Louchart et al., Phys. Rev. C 87, 054302 (2013).
\bibitem{Lenzi2010}
S. M. Lenzi, F. Nowacki, A. Poves, and K. Sieja, Phys. Rev. 82, 054301 (2010).

\end{thebibliography}

\clearpage

\appendix
\section{}

\begingroup
\setlength{\tabcolsep}{6pt}
\renewcommand{\arraystretch}{1.3}
\LTcapwidth=0.45\textwidth
\label{appendix:en}
\begin{longtable}{|c  c  c  c c|}
\caption{Comparison between calculated and experimental low-lying excitation energies for the $^{18}$O, $^{17}$O, $^{76}$Se and $^{77}$Se nuclei. Energies are in MeV.}
\label{table:teor_en}
\endfirsthead
\hline
Nucleus & J$^{\pi}$ & E$_{EXP.}$ & E$_{SM}$ & E$_{IBFM}$\\
\hline
$^{18}$O & 0$^{+}$ & 0 & 0 & -\\
 & 2$^{+}$ & 1.982 & 2.045 & -\\
 \hline
$^{17}$O & 5/2$^{+}$ & 0 & 0 & -\\
 & 1/2$^{+}$ & 0.870 & 0.957 & -\\
\hline
$^{76}$Se & 0$^{+}$ & 0 & 0 & \\
 & 2$^{+}$ & 0.559 & 0.722 & \\
 \hline
 & & ROI 1 & &\\
  & 1/2$^{-}$ & 0 & 0.584 & 0\\
  & 7/2$^{+}$ & 0.161 & 0.141 & -\\
 $^{77}$Se & 9/2$^{+}$ & 0.175 & 0 & 1.726\\
  & 3/2$^{-}$ & 0.238 & 0.953 & 0.200\\
  & 5/2$^{-}$ & 0.249 & 0.441 & 0.211\\
  & 5/2$^{+}$ & 0.301 & 0.302 & -\\
  & & & &\\
  & & ROI 2 & &\\
  & 5/2$^{-}$ & 0.439 & 1.190 & 0.686\\
  & 3/2$^{-}$ & 0.520 & 1.220 & 0.508\\
  & 7/2$^{-}$ & 0.581 & 1.227 & -\\
$^{77}$Se  & 5/2$^{+}$ & 0.680 & 0.784 & -\\
  & 7/2$^{+}$ & 0.796 & 1.269 & -\\
  & 7/2$^{-}$ & 0.808 & 1.787 & -\\
  & 1/2$^{-}$ & 0.817 & 1.087 & 0.930\\
  & 5/2$^{-}$ & 0.824 & 1.437 & 1.152\\
  & & & &\\
  & & ROI 3 & &\\
  & 11/2$^{+}$ & 0.970 & 0.835 & -\\
  & 9/2$^{-}$ & 0.978 & 1.238 & -\\
  & 3/2$^{-}$ & 1.005 & 1.651 & 1.116\\
  & 13/2$^{+}$ & 1.024 & 0.738 & -\\
  & 11/2$^{+}$ & 1.126 & 1.337 & -\\
  & 9/2$^{-}$ & 1.172 & 1.941 & -\\
 $^{77}$Se & 5/2$^{-}$ & 1.179 & 1.803 & 1.936\\
  & 3/2$^{-}$ & 1.186 & 1.890 & 1.284\\
  & 9/2$^{+}$ & 1.193 & 1.123 & 2.342\\
  & 5/2$^{-}$ & 1.230 & 1.896 & 1.952\\
  & 5/2$^{+}$ & 1.252 & 1.710 & -\\
  & 7/2$^{-}$ & 1.282 & 2.194 & -\\
  & 3/2$^{-}$ & 1.364 & 2.157 & 2.289\\
  & 5/2$^{-}$ & 1.529 & 2.357 & -\\
  & 5/2$^{+}$ & 1.607 & 2.052 & -\\
  & 1/2$^{-}$ & 1.623 & 1.743 & 2.291\\
  \hline

\end{longtable}
\endgroup

\clearpage

\begingroup
\setlength{\tabcolsep}{6pt}
\renewcommand{\arraystretch}{1.3}
\LTcapwidth=0.45\textwidth

\label{appendix:SA}
\begin{longtable}{|c  c  c  c c|}
\caption{Shell Model (SM) spectroscopic amplitudes (SA) used in DWBA and CCBA calculations. Interacting boson-fermion model (IBFM-2) spectroscopic amplitudes (SA) used in DWBA calculations.}
\label{table:SA_KSHELL}
\endfirsthead
\hline
Initial State & $nlj$ & Final State & SA$_{SM}$ & SA$_{IBFM}$\\ 
\hline
$^{18}$O$_{g.s.}$(0$^{+}$) & $1d_{5/2}$ & $^{17}$O$_{g.s.}$(5/2$^{+}$)  & 1.3039 & -\\
 & $2s_{1/2}$ & $^{17}$O$_{0.870}$(1/2$^{+}$)  & -0.5606 & -\\
\hline
 & $1d_{5/2}$ & $^{17}$O$_{g.s.}$(5/2$^{+}$)  & -0.9283 & -\\
$^{18}$O$_{1.982}$(2$^{+}$) & $2s_{1/2}$ & $^{17}$O$_{g.s.}$(5/2$^{+}$)  & -0.6661 & -\\
& $1d_{5/2}$ & $^{17}$O$_{0.870}$(1/2$^{+}$)  & 0.6514 & -\\
\hline
 & & ROI 1 & &\\
 & $1p_{1/2}$ & $^{77}$Se$_{g.s.}$(1/2$^{-}$)  & -0.4113 & 0.1566\\
$^{76}$Se$_{g.s.}$(0$^{+}$) & $0g_{9/2}$ & $^{77}$Se$_{0.175}$(9/2$^{+}$)  & -0.6916 & 0.3463\\
 & $1p_{3/2}$ & $^{77}$Se$_{0.238}$(3/2$^{-}$)  & 0.0401 & 0.2044\\
 & $0f_{5/2}$ & $^{77}$Se$_{0.249}$(5/2$^{-}$)  & -0.5294 & 0.2348\\
 & & & &\\
 & & ROI 2 & &\\
 & $0f_{5/2}$ & $^{77}$Se$_{0.439}$(5/2$^{-}$)  & -0.0029 & -0.0221\\
$^{76}$Se$_{g.s.}$(0$^{+}$) & $1p_{3/2}$ & $^{77}$Se$_{0.520}$(3/2$^{-}$)  & 0.2449 & -0.1901\\
 & $1p_{1/2}$ & $^{77}$Se$_{0.817}$(1/2$^{-}$)  & 0.1097 & -0.0101\\
 & $0f_{5/2}$ & $^{77}$Se$_{0.824}$(5/2$^{-}$)  & 0.0120 & -0.4967\\
 & & & &\\
 & & ROI 3 & &\\
 & $1p_{3/2}$ & $^{77}$Se$_{1.005}$(3/2$^{-}$)  & 0.0232 & 0.2646\\
 & $0f_{5/2}$ & $^{77}$Se$_{1.179}$(5/2$^{-}$)  & 0.0291 & -0.3385\\
 & $1p_{3/2}$ & $^{77}$Se$_{1.186}$(3/2$^{-}$)  & 0.0071 & 0.0004\\
$^{76}$Se$_{g.s.}$(0$^{+}$) & $0g_{9/2}$ & $^{77}$Se$_{1.193}$(9/2$^{+}$)  & -0.0401 & -0.3192\\
 & $0f_{5/2}$ & $^{77}$Se$_{1.230}$(5/2$^{-}$)  & 0.0541 & -0.3040\\
 & $1p_{3/2}$ & $^{77}$Se$_{1.364}$(3/2$^{-}$)  & 0.0502 & 0.0681\\
 & $0f_{5/2}$ & $^{77}$Se$_{1.529}$(5/2$^{-}$)  & 0.0306 & -\\
 & $1p_{1/2}$ & $^{77}$Se$_{1.623}$(1/2$^{-}$)  & -0.0455 & 0.1648\\
 \hline
 & & ROI 1 & &\\
 & $1p_{3/2}$ & $^{77}$Se$_{g.s.}$(1/2$^{-}$)  & -0.3605 & -\\
 & $0f_{5/2}$ & $^{77}$Se$_{g.s.}$(1/2$^{-}$) & -0.4787 & -\\
 & $0g_{9/2}$ & $^{77}$Se$_{0.161}$(7/2$^{+}$)  & -0.7549 & -\\
 & $0g_{9/2}$ & $^{77}$Se$_{0.175}$(9/2$^{+}$)  & -0.4447 & -\\
$^{76}$Se$_{0.559}$(2$^{+}$) & $1p_{1/2}$ & $^{77}$Se$_{0.238}$(3/2$^{-}$) & -0.0614 & -\\
 & $1p_{3/2}$ & $^{77}$Se$_{0.238}$(3/2$^{-}$)  & 0.0717 & -\\
 & $0f_{5/2}$ & $^{77}$Se$_{0.238}$(3/2$^{-}$) & 0.3690 & -\\
 & $1p_{1/2}$ & $^{77}$Se$_{0.249}$(5/2$^{-}$) & -0.3111 & -\\
 & $1p_{3/2}$ & $^{77}$Se$_{0.249}$(5/2$^{-}$)  & -0.0963 & -\\
 & $0f_{5/2}$ & $^{77}$Se$_{0.249}$(5/2$^{-}$) & -0.5023 & -\\
 & $0g_{9/2}$ & $^{77}$Se$_{0.301}$(5/2$^{+}$)  & -0.6450 & -\\
 & & & &\\
 & & ROI 2 & &\\
 & $1p_{1/2}$ & $^{77}$Se$_{0.439}$(5/2$^{-}$) & -0.2528 & -\\
 & $1p_{3/2}$ & $^{77}$Se$_{0.439}$(5/2$^{-}$)  & -0.0584 & -\\
 & $0f_{5/2}$ & $^{77}$Se$_{0.439}$(5/2$^{-}$) & 0.0502 & -\\
 & $1p_{1/2}$ & $^{77}$Se$_{0.520}$(3/2$^{-}$) & -0.3671 & -\\
  \hline
 Initial State & $nlj$ & Final State & SA$_{SM}$ & SA$_{IBFM}$\\
 \hline
 & $1p_{3/2}$ & $^{77}$Se$_{0.520}$(3/2$^{-}$)  & 0.2143 & -\\
 & $0f_{5/2}$ & $^{77}$Se$_{0.520}$(3/2$^{-}$) & -0.2596 & -\\
$^{76}$Se$_{0.559}$(2$^{+}$) & $1p_{3/2}$ & $^{77}$Se$_{0.581}$(7/2$^{-}$)  & 0.0057 & -\\
 & $0f_{5/2}$ & $^{77}$Se$_{0.581}$(7/2$^{-}$) & 0.3163 & -\\
 & $0g_{9/2}$ & $^{77}$Se$_{0.680}$(5/2$^{+}$)  & -0.4774 & -\\
 & $0g_{9/2}$ & $^{77}$Se$_{0.796}$(7/2$^{+}$)  & 0.0736 & -\\
 & $1p_{3/2}$ & $^{77}$Se$_{0.808}$(7/2$^{-}$)  & 0.0550 & -\\
 & $0f_{5/2}$ & $^{77}$Se$_{0.808}$(7/2$^{-}$) & -0.0225 & -\\
 & $1p_{3/2}$ & $^{77}$Se$_{0.817}$(1/2$^{-}$)  & 0.0243 & -\\
 & $0f_{5/2}$ & $^{77}$Se$_{0.817}$(1/2$^{-}$) & 0.6685 & -\\
 & $1p_{1/2}$ & $^{77}$Se$_{0.824}$(5/2$^{-}$) & 0.1957 & -\\
 & $1p_{3/2}$ & $^{77}$Se$_{0.824}$(5/2$^{-}$)  & 0.0229 & -\\
 & $0f_{5/2}$ & $^{77}$Se$_{0.824}$(5/2$^{-}$) & 0.2239 & -\\
 & & & &\\
 & & ROI 3 & &\\
 & $0g_{9/2}$ & $^{77}$Se$_{0.970}$(11/2$^{+}$)  & -0.4719 & -\\
 & $0f_{5/2}$ & $^{77}$Se$_{0.978}$(9/2$^{-}$)  & 0.5210 & -\\
 & $1p_{1/2}$ & $^{77}$Se$_{1.005}$(3/2$^{-}$) & 0.1451 & -\\
 & $1p_{3/2}$ & $^{77}$Se$_{1.005}$(3/2$^{-}$)  & 0.0672 & -\\
 & $0f_{5/2}$ & $^{77}$Se$_{1.005}$(3/2$^{-}$) & 0.1507 & -\\
 & $0g_{9/2}$ & $^{77}$Se$_{1.024}$(13/2$^{+}$)  & -0.7689 & -\\
 & $0g_{9/2}$ & $^{77}$Se$_{1.126}$(11/2$^{+}$)  & -0.2260 & -\\
 & $0f_{5/2}$ & $^{77}$Se$_{1.172}$(9/2$^{-}$)  & 0.0677 & -\\
 & $1p_{1/2}$ & $^{77}$Se$_{1.179}$(5/2$^{-}$) & 0.0833 & -\\
 & $1p_{3/2}$ & $^{77}$Se$_{1.179}$(5/2$^{-}$)  & 0.0869 & -\\
 & $0f_{5/2}$ & $^{77}$Se$_{1.179}$(5/2$^{-}$) & 0.0490 & -\\
 & $1p_{1/2}$ & $^{77}$Se$_{1.186}$(3/2$^{-}$) & -0.1499 & -\\
& $1p_{3/2}$ & $^{77}$Se$_{1.186}$(3/2$^{-}$)  & 0.1241 & -\\
$^{76}$Se$_{0.559}$(2$^{+}$) & $0f_{5/2}$ & $^{77}$Se$_{1.186}$(3/2$^{-}$) & 0.0063 & -\\
 & $0g_{9/2}$ & $^{77}$Se$_{1.193}$(9/2$^{+}$)  & -0.2213 & -\\
 & $0f_{5/2}$ & $^{77}$Se$_{1.230}$(5/2$^{-}$) & 0.0580 & -\\
 & $0f_{5/2}$ & $^{77}$Se$_{1.230}$(5/2$^{-}$) & -0.0475 & -\\
 & $0f_{5/2}$ & $^{77}$Se$_{1.230}$(5/2$^{-}$) & 0.1143 & -\\
  & $0g_{9/2}$ & $^{77}$Se$_{1.252}$(5/2$^{+}$)  & 0.3737 & -\\
 & $1p_{3/2}$ & $^{77}$Se$_{1.282}$(7/2$^{-}$) & 0.0920 & -\\
 & $0f_{5/2}$ & $^{77}$Se$_{1.282}$(7/2$^{-}$) & -0.0848 & -\\
 & $1p_{1/2}$ & $^{77}$Se$_{1.364}$(3/2$^{-}$) & -0.0969 & -\\
 & $1p_{3/2}$ & $^{77}$Se$_{1.364}$(3/2$^{-}$) & 0.0704 & -\\
 & $0f_{5/2}$ & $^{77}$Se$_{1.364}$(3/2$^{-}$) & -0.0024 & -\\
 & $1p_{1/2}$ & $^{77}$Se$_{1.529}$(5/2$^{-}$) & -0.0634 & -\\
 & $1p_{3/2}$ & $^{77}$Se$_{1.529}$(5/2$^{-}$) & -0.0091 & -\\
 & $0f_{5/2}$ & $^{77}$Se$_{1.529}$(5/2$^{-}$) & -0.0740 & -\\
 & $0g_{9/2}$ & $^{77}$Se$_{1.607}$(5/2$^{+}$)  & -0.0439 & -\\
 & $1p_{3/2}$ & $^{77}$Se$_{1.623}$(1/2$^{-}$) & -0.1692 & -\\
 & $0f_{5/2}$ & $^{77}$Se$_{1.623}$(1/2$^{-}$) & -0.0053 & -\\
 \hline
\end{longtable}
\endgroup

\clearpage

\begingroup
\setlength{\tabcolsep}{6pt}
\renewcommand{\arraystretch}{1.3}
\LTcapwidth=0.45\textwidth
\label{appendix:ME2}
\begin{longtable}{|c  c  c  c c c c c|}
\caption{Adopted reduced matrix elements M(E2) and deformation lengths $\delta_{2}$ in the final partition.}
\label{table:ME2}
\endfirsthead
\hline
Nucleus & Transition & Initial state & Final state & M(E2) & $\delta_{2}$ & B$_{th}$(E2) & B$_{exp}$(E2) \\
 & $J^{\pi} \longleftrightarrow J^{'\pi}$ & (MeV) & (MeV) & (e fm$^2$) & (fm) & W.u. & W.u.\\
\hline
$^{17}$O & $5/2^{+} \longleftrightarrow 1/2^{+}$ & 0 & 0.870 & 3.77 & 0.42 & 2.74 & 2.39\\
\hline
 & $1/2^{-} \longleftrightarrow 5/2^{-}$ & 0.000 & 0.249 & 26.25 & 0.69 & 5.90 & 1.98 (5)\\
 & $1/2^{-} \longleftrightarrow 3/2^{-}$ & 0.000 & 0.520 & 43.63 & 1.14 & 24.46 & 42.2 (25)\\
 & $7/2^{+} \longleftrightarrow 9/2^{+}$ & 0.161 & 0.175 & -95.50 & -2.50 & 46.87	& -\\
 & $9/2^{+} \longleftrightarrow 11/2^{+}$ & 0.175 & 0.970 & 69.13 & 1.81 & 20.47 & -\\
$^{77}$Se & $9/2^{+} \longleftrightarrow 13/2^{+}$ & 0.175 & 1.024 & 109.50 & 2.87 & 44.01 & 1.9 E2 (8)\\
 & $9/2^{+} \longleftrightarrow 5/2^{+}$ & 0.175 & 1.252 & 50.26 & 1.32 & 21.64 & -\\
 & $5/2^{-} \longleftrightarrow 3/2^{-}$ & 0.249 & 0.520 & -27.67 & -0.73 & 9.84 & 5.3 (+57 -25)\\
 & $5/2^{-} \longleftrightarrow 7/2^{-}$ & 0.249 & 0.581 & -63.18 & -1.66 & 25.64	& 74 (23)\\
 & $5/2^{-} \longleftrightarrow 1/2^{-}$ & 0.249 & 0.817 & -33.41 & -0.88 & 28.68	& -\\
 & $5/2^{-} \longleftrightarrow 9/2^{-}$ & 0.249 & 0.978 & -79.37 & -2.08 & 32.37	& 1.3 E2 (7)\\
  \hline
\end{longtable}
\endgroup

\clearpage

\begingroup
\setlength{\tabcolsep}{6pt}
\renewcommand{\arraystretch}{1.3}
\LTcapwidth=0.45\textwidth
\label{appendix:cross}
\begin{longtable}{|c  c  c  c|}
\caption{One-neutron transfer theoretical cross sections (in $\mu b$) integrated in the angular range [4$\deg $- 12.5$\deg$] of the laboratory system and for all the combination of projectile and target states lying within the three ROI in Fig. \ref{fig:xsec_fit}.}
\label{table:integrated_cross_sec}
\endfirsthead
\hline
Final & DWBA & CCBA & DWBA \\
Channel & SM & SM & IBFM\\
\hline
& ROI 1 & & \\
$^{17}$O$_{g.s.}$(5/2$^{+}$)+$^{77}$Se$_{g.s.}$(1/2$^{-}$)  & 109.49 & 172.55 & 15.87\\
$^{17}$O$_{g.s.}$(5/2$^{+}$)+$^{77}$Se$_{0.161}$(7/2$^{+}$) & - & 22.00 & - \\
$^{17}$O$_{g.s.}$(5/2$^{+}$)+$^{77}$Se$_{0.175}$(9/2$^{+}$) & 1316.63 & 1591.79 & 330.11\\
$^{17}$O$_{g.s.}$(5/2$^{+}$)+$^{77}$Se$_{0.238}$(3/2$^{-}$) & 3.28 & 5.53 & 85.33\\
$^{17}$O$_{g.s.}$(5/2$^{+}$)+$^{77}$Se$_{0.249}$(5/2$^{-}$) & 447.74 & 634.77 & 88.08\\
$^{17}$O$_{g.s.}$(5/2$^{+}$)+$^{77}$Se$_{0.301}$(5/2$^{+}$) & - & 73.43 & - \\
& & & \\
& ROI 2 & & \\
$^{17}$O$_{g.s.}$(5/2$^{+}$)+$^{77}$Se$_{0.439}$(5/2$^{-}$) & 0.01 & 5.24 & 0.79\\
$^{17}$O$_{g.s.}$(5/2$^{+}$)+$^{77}$Se$_{0.520}$(3/2$^{-}$) & 118.79 & 117.53 & 71.58\\
$^{17}$O$_{g.s.}$(5/2$^{+}$)+$^{77}$Se$_{0.581}$(7/2$^{-}$) & - & 8.79 & -\\
$^{17}$O$_{g.s.}$(5/2$^{+}$)+$^{77}$Se$_{0.680}$(5/2$^{+}$) & - & 54.20 & - \\
$^{17}$O$_{g.s.}$(5/2$^{+}$)+$^{77}$Se$_{0.796}$(7/2$^{+}$) & - & 0.04 & - \\
$^{17}$O$_{g.s.}$(5/2$^{+}$)+$^{77}$Se$_{0.808}$(7/2$^{-}$) & - & 0.08 & - \\
$^{17}$O$_{g.s.}$(5/2$^{+}$)+$^{77}$Se$_{0.817}$(1/2$^{-}$) & 7.07 & 19.93 & 0.06\\
$^{17}$O$_{g.s.}$(5/2$^{+}$)+$^{77}$Se$_{0.824}$(5/2$^{-}$) & 0.17 & 5.59 & 411.47\\
$^{17}$O$_{0.870}$(1/2$^{+}$)+$^{77}$Se$_{g.s.}$(1/2$^{-}$) & 55.21 & 83.29 & 8.00\\
& & & \\
& ROI 3 & & \\
$^{17}$O$_{g.s.}$(5/2$^{+}$)+$^{77}$Se$_{0.970}$(11/2$^{+}$) & - & 7.17 & - \\
$^{17}$O$_{g.s.}$(5/2$^{+}$)+$^{77}$Se$_{0.978}$(9/2$^{-}$) & - & 45.31 & - \\
$^{17}$O$_{g.s.}$(5/2$^{+}$)+$^{77}$Se$_{1.005}$(3/2$^{-}$) & 1.00 & 12.36 & 130.60\\
$^{17}$O$_{g.s.}$(5/2$^{+}$)+$^{77}$Se$_{1.024}$(13/2$^{+}$ & - & 83.83 & - \\
$^{17}$O$_{0.870}$(1/2$^{+}$))+$^{77}$Se$_{0.161}$(7/2$^{+}$) & - & 5.55 & - \\
$^{17}$O$_{0.870}$(1/2$^{+}$))+$^{77}$Se$_{0.175}$(9/2$^{+}$) & 377.72 & 370.20 & 94.70\\
$^{17}$O$_{0.870}$(1/2$^{+}$)+$^{77}$Se$_{0.238}$(3/2$^{-}$) & 1.05 & 1.79 & 27.33\\
$^{17}$O$_{0.870}$(1/2$^{+}$)+$^{77}$Se$_{0.249}$(5/2$^{-}$) & 191.11 & 232.12 & 37.59\\
$^{17}$O$_{g.s.}$(5/2$^{+}$)+$^{77}$Se$_{1.126}$(11/2$^{+}$) & - & 0.55 & - \\
$^{17}$O$_{0.870}$(1/2$^{+}$)+$^{77}$Se$_{0.301}$(5/2$^{+}$) & - & 14.08 & - \\
$^{17}$O$_{g.s.}$(5/2$^{+}$)+$^{77}$Se$_{1.172}$(9/2$^{-}$) & - & 0.22 & - \\
$^{17}$O$_{g.s.}$(5/2$^{+}$)+$^{77}$Se$_{1.179}$(5/2$^{-}$) & 1.45 & 4.17 & 195.79\\
$^{17}$O$_{g.s.}$(5/2$^{+}$)+$^{77}$Se$_{1.186}$(3/2$^{-}$) & 0.09 & 1.11 & 0.00\\
$^{17}$O$_{g.s.}$(5/2$^{+}$)+$^{77}$Se$_{1.193}$(9/2$^{+}$) & 4.97 & 10.71 & 314.90\\
$^{17}$O$_{g.s.}$(5/2$^{+}$)+$^{77}$Se$_{1.230}$(5/2$^{-}$) & 5.02 & 1.68 & 158.46\\
$^{17}$O$_{g.s.}$(5/2$^{+}$)+$^{77}$Se$_{1.252}$(5/2$^{+}$) & - & 6.36 & - \\
$^{17}$O$_{g.s.}$(5/2$^{+}$)+$^{77}$Se$_{1.282}$(7/2$^{-}$) & - & 1.44 & - \\
$^{17}$O$_{0.870}$(1/2$^{+}$)+$^{77}$Se$_{0.439}$(5/2$^{-}$) & 0.01 & 1.27 & 0.34\\
$^{17}$O$_{g.s.}$(5/2$^{+}$)+$^{77}$Se$_{1.364}$(3/2$^{-}$) & 0.17 & 0.82 & 8.22\\
$^{17}$O$_{0.870}$(1/2$^{+}$)+$^{77}$Se$_{0.520}$(3/2$^{-}$) & 39.23 & 33.55 & 25.52\\
$^{17}$O$_{0.870}$(1/2$^{+}$)+$^{77}$Se$_{0.581}$(7/2$^{-}$) & - & 2.63 & - \\
$^{17}$O$_{g.s.}$(5/2$^{+}$)+$^{77}$Se$_{1.529}$(5/2$^{-}$) & 1.62 & 0.16 & -\\
$^{17}$O$_{0.870}$(1/2$^{+}$)+$^{77}$Se$_{0.680}$(5/2$^{+}$) & - & 11.06 & - \\
\hline
Final & DWBA & CCBA & DWBA \\
Channel & SM & SM & IBFM\\
\hline
$^{17}$O$_{g.s.}$(5/2$^{+}$)+$^{77}$Se$_{1.607}$(5/2$^{+}$) & - & 0.18 & - \\
$^{17}$O$_{g.s.}$(5/2$^{+}$)+$^{77}$Se$_{1.623}$(1/2$^{-}$) & 1.08 & 1.67 & 14.20\\
$^{17}$O$_{0.870}$(1/2$^{+}$)+$^{77}$Se$_{0.796}$(7/2$^{+}$) & - & 0.01 & - \\
$^{17}$O$_{0.870}$(1/2$^{+}$)+$^{77}$Se$_{0.808}$(7/2$^{-}$) & - & 0.02 & - \\
$^{17}$O$_{0.870}$(1/2$^{+}$)+$^{77}$Se$_{0.817}$(1/2$^{-}$) & 2.93 & 0.86 & 2.36\\
$^{17}$O$_{0.870}$(1/2$^{+}$)+$^{77}$Se$_{0.824}$(5/2$^{-}$) & 0.11 & 1.68 & 187.29\\
 \hline
\end{longtable}
\endgroup

\end{document}